\documentclass[12pt]{article}
\usepackage{graphicx}
\input epsf.tex

\let\badcite=\cite
\def\cite{~\badcite}

\def\slashchar#1{\setbox0=\hbox{$#1$}           
   \dimen0=\wd0                                 
   \setbox1=\hbox{/} \dimen1=\wd1               
   \ifdim\dimen0>\dimen1                        
      \rlap{\hbox to \dimen0{\hfil/\hfil}}      
      #1                                        
   \else                                        
      \rlap{\hbox to \dimen1{\hfil$#1$\hfil}}   
      /                                         
   \fi} 
    \def\slashword#1{\setbox0=\hbox{$#1$}        
  \dimen0=\wd0                                   
   \setbox1=\hbox{/} \dimen1=\wd1                
   \ifdim\dimen0>\dimen1                         
      \rlap{\hbox to \dimen0{\hfil\bf---\hfil}} %
      #1                                         %
   \else                                         
      \rlap{\hbox to \dimen1{\hfil$#1$\hfil}}    
      /                                          
    \fi}                                         %

\catcode`@=11
\newdimen\vbigd@men                             

\def\vbig#1#2{{\vbigd@men=#2\divide\vbigd@men by 2%
   \hbox{$\left#1\vbox to \vbigd@men{}\right.\n@space$}}}
\catcode`@=12

\catcode`@=11
\def\citenum#1{\csname b@#1\endcsname}
\catcode`@=12

\def\lsim{\raisebox{-.1em}{$
\buildrel{\scriptscriptstyle <}\over{\scriptscriptstyle\sim}$}}

\def\dofig#1#2{\centerline{\epsfxsize=#1\epsfbox{#2}}}
\begin{document}
\begin{titlepage}

\begin{flushright}
{SCUPHY-TH-06003}\\
{TUHEP-TH-06155}\\   
\end{flushright}

\bigskip\bigskip

\begin{center}{\Large\bf\boldmath
Wedgebox analysis of four-lepton events from neutralino pair production 
at the LHC}
\end{center}
\bigskip
\centerline{\bf G. Bian, M. Bisset}
\centerline{{\it Center for High Energy Physics and Department of
Physics,}}
\centerline{{\it Tsinghua University, P.R. China}}
\centerline{\bf N. Kersting, Y. Liu, X. Wang}
\centerline{{\it Physics Department, Sichuan University, P.R. China}}

\bigskip

\begin{abstract}

A `wedgebox' plot is a two-dimensional scatter-plot of two invariant mass
quantities.  Here $pp \to e^+e^-\mu^+\mu^- + \slashchar{E}$
signature LHC events are analyzed by plotting the di-electron 
invariant mass {\it versus} the di-muon invariant mass.  Data 
sets of such events are obtained across the MSSM input parameter 
space in realistic event-generator simulations, including cuts designed 
to remove SM backgrounds.  Their study reveals several general features.  
Firstly, regions in the MSSM input parameter space where a sufficient 
number of events are expected so as to be able to construct a clear 
wedgebox plot are delineated. 
Secondly, the presence of box shapes on a wedgebox plot either indicates 
the presence of heavy Higgs bosons decays or restricts the location to a 
quite small region of low $\mu$ and $M_2$ values 
$\lsim\, 200\, \hbox{GeV}$, a region denoted as the `lower island'.  
In this region, wedgebox plots can be quite complicated and change in 
pattern rather quickly as one moves around in the $( \mu , M_2 )$ plane. 
Thirdly, direct neutralino pair production from an intermediate $Z^{0*}$ 
may only produce a wedge-shape since only 
$\widetilde{\chi}_2^0\widetilde{\chi}_3^0$ decays can contribute 
significantly.  And fourthly, a double-wedge or 
wedge-protruding-from-a-box pattern on a wedgebox plot, which results 
from combining a variety of MSSM production processes, yields three 
distinct observed endpoints, almost always attributable to  
$\widetilde{\chi}_{2,3,4}^0 \to \widetilde{\chi}_1^0  \ell^+\ell^-$
decays, which can be utilized to determine a great deal of information
about the neutralino and slepton mass spectra and related MSSM input 
parameters.  Wedge and double-wedge patterns are seen in wedgebox plots
in another region of higher $\mu$ and $M_2$ values, denoted as the 
`upper island.'  Here the pattern is simpler and more stable as one
moves across the $( \mu , M_2 )$ input parameter space.

\bigskip        

\end{abstract}

\newpage
\pagestyle{empty}

\end{titlepage}


\section{Introduction}

The Large Hadron Collider (LHC) is scheduled to begin operation in 
less than two years, at which time the predictions of models of particle 
physics beyond the Standard Model (SM), especially
Supersymmetry (SUSY), will be confronted with significant, and 
potentially lethal, experimental constraints.
SUSY predicts heavy scalar counterparts, or superpartners, to the SM 
fermions, as well as fermionic superpartners to the SM bosons ---
both the spin-1 gauge bosons and the spin-0 Higgs bosons (SUSY 
requires more than one Higgs boson).  
These new states are known collectively as sparticles.
Colorless sparticles,  including the neutralinos ($\widetilde{\chi}_i^0$) 
and charginos ($\widetilde{\chi}_j^{\pm}$) --- the neutral and charged, 
respectively, superpartners of admixtures of the gauge and Higgs bosons --- 
are generally expected to be somewhat lighter than their colored brethren, 
the gluinos ($\tilde{g}$) and squarks ($\tilde{q}$).  Nonetheless these 
latter are expected to have the largest production 
cross-sections, unless they are an order of magnitude or more massive.  
Yet this is precisely what 
occurs in some hypothesized SUSY-breaking scenarios: the squarks and 
gluinos have masses on the scale of several TeV while neutralinos and 
charginos have masses on the scale of several hundred GeV (or less). 
Therefore a study of these sparticles' direct production modes is called 
for.
Moreover, 
aside from their direct production modes, colorless sparticles inevitably
appear {\em indirectly} in any colored sparticle production process through 
the sometimes complicated decay chains of the gluinos and squarks.
Thus, determining the masses and couplings of the neutralinos and charginos 
is crucial to understanding almost any SUSY events which may emerge at the 
LHC.

\begin{figure}[t!]
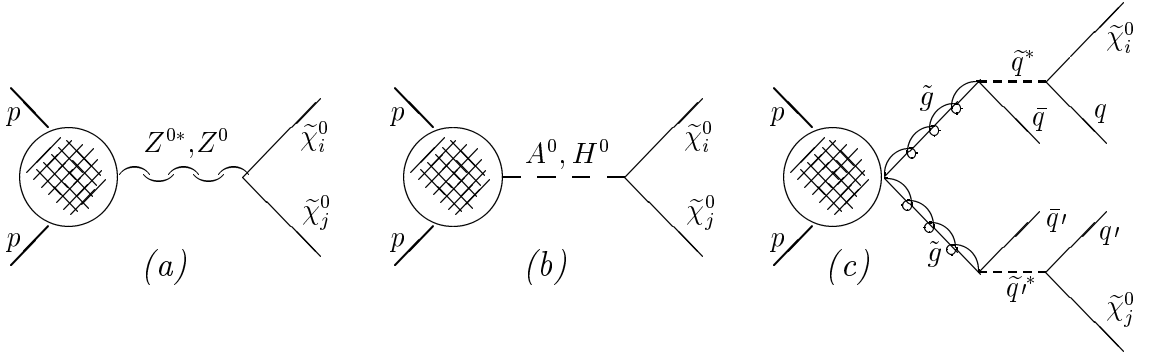

\vskip 1.2cm
\dofig{6.10in}{figure1.eps}
\vskip -18.7cm
\caption{{\small
Feynman diagrams for heavy ($i,j = 2,3,4$) neutralino pair production
\newline
mechanisms:
{\it (a)} `direct' production via EW gauge boson;
{\it (b)} Higgs-mediated production; and
{\it (c)} production via cascade decays of gluinos (shown here)
or squarks (make squarks in diagram on-mass shell and 
remove the gluinos and the connected quarks).  
}}
\label{inoinoprod}
\end{figure}

In the R-parity-conserving Minimal Supersymmetric Standard Model (MSSM),
sparticles must be pair-produced, and the lightest sparticle (the LSP), 
for which the preferred candidate is generally the lightest neutralino
($\widetilde{\chi}_1^0$), is stable.  
The focus of this study is the neutralinos, of which there are four in the 
MSSM, and in particular the 
heavier three ($\widetilde{\chi}_i^0$, $i=2,3,4$ in order of increasing 
mass) --- which are expected to decay, either directly or indirectly, 
into the LSP.
MSSM neutralino pair production at the LHC can in general occur via 
three avenues herein known as: 
direct, Higgs-mediated, and colored-sparticle cascade decays, as shown 
in Fig.\ \ref{inoinoprod}.  Cascade decays were studied in 
\cite{cascade}, while \cite{ha4l} focused on Higgs-mediated decays.
The present study enlarges the focus of \cite{ha4l} to also encompass
the direct production channel via the electroweak (EW) $Z^0$ gauge 
boson, which formed an unavoidable and often significant background in the 
\cite{ha4l} study.  
This direct avenue is most dominant when the colored sparticles and 
the extra Higgs bosons of the MSSM are quite massive (such as if squark, 
gluino, and pseudoscalar Higgs MSSM input masses are set around the TeV scale).

In an LHC detector, each short-lived heavier neutralino produced must 
decay into an LSP and SM particles (with the invisible LSP, and any SM 
neutrinos that may be present, generating the tell-tale SUSY missing 
energy signature).  Rates for the resulting final state combinations of
observed SM particles, and the distributions of the energies and momenta
of said SM particles, will depend on MSSM (especially neutralino) masses 
and couplings.  It would be simplest to examine final states that are 
produced by only one unique pair of neutralinos, and, on top of this, 
via only one of the aforementioned neutralino pair production avenues.
Clearly though this is not a realistic expectation.

In the present study, as well as in \cite{cascade}, the signature examined 
is neutralino pair decays into an electron-positron pair, a muon-antimuon 
pair, and missing energy (and possibly jets):
 $ pp \to \widetilde{\chi}_i^0 \widetilde{\chi}_j^0 \to e^+e^-   
\mu^+\mu^- \, + \, \slashchar{E} \, ( + n\, \rm{jet}) $,
where all leptons are hard and isolated (exact conditions for these
requirements will be given later).
The rationales for choosing this particular final state are two-fold: 
first, 
in the hadronically noisy environment of the LHC, multi-lepton signals 
have minimal SM backgrounds and thus tend to be easier to identify. 
Second,
assuming the neutralinos proceed to this final state via one of the 
following decay chains,
\begin{eqnarray}
 & \widetilde{\chi}_i^0 \to \{ Z^0,Z^{0*} \} + \widetilde{\chi}_1^0 
\to \ell^+\ell^- + \widetilde{\chi}_1^0 & 
\label{basic-decaysZ}
\\
\hbox{or} \;\;\; 
 & \widetilde{\chi}_i^0 
\to \ell^\mp + \{ \tilde{\ell}^\pm,\tilde{\ell}^{\pm *} \}  
\to \ell^+\ell^- + \widetilde{\chi}_1^0 & \;\;\; , 
\label{basic-decaysL}
\end{eqnarray}
where ${\ell} = e \, \hbox{or}\, \mu$, the dilepton invariant masses are 
cleanly bounded by
\begin{eqnarray}
& 0 < & M_{i1}(\ell^+\ell^-) \;\; < 
m_{\widetilde{\chi}_i^0} - m_{\widetilde{\chi}_1^0}  
\label{edge3body}
\\
 \nonumber \hbox{or} \!\!\!\!\!\!\!\!\!\!\!\!\! & \\
& 0 < & M_{i1}(\ell^+\ell^-) \;\; < m_{\widetilde{\chi}_i^0} 
\sqrt{ 1 - \left(\frac{m_{\tilde{\ell}}}
                      {m_{\widetilde{\chi}_i^0}}\right)^{\!\!\! 2}}
\sqrt{ 1 - 
\left(\frac{m_{\widetilde{\chi}_1^0}}
           {m_{\tilde{\ell}}}\right)^{\!\!\! 2}} \,\,\, , 
\label{edge2body}
\end{eqnarray}
depending on whether the decays are 3-body (via $Z^{0*}$ or
$\tilde{\ell}^{\pm *}$) or 2-body via an on-mass-shell charged slepton, 
respectively.  A 2-body decay via an on-mass-shell $Z^0$ leads to  
$M_{i1}(\ell^+\ell^-) = M_Z$, which is non-trivial to extract from
SM backgrounds. 
The fact that the dilepton invariant mass spectrum basically increases as 
one runs up in mass to the upper kinematical edge \cite{PaigeII} 
greatly facilitates a precise determination of this bound.  
Then, if the electron and muon pairs always come from one particular 
${ij}$-combination of neutralinos, plotting their dilepton invariant 
masses against each other in a two-dimensional $M(e^+e^-)$ 
versus $M(\mu^+\mu^-)$ Dalitz-like plot \cite{cascade} will yield either 
a box- (for $i=j$) or wedge-shape (for $i \ne j$) with hard kinematical 
edges at (\ref{edge3body}) or (\ref{edge2body}).
Note that here the lepton pairs are required to be of different flavors to 
facilitate proper pairings.

However, the situation is complicated at the LHC (where the partonic 
center-of-mass is not fixed) by the fact that several different 
${ij}$-combinations may be produced --- each at a different rate.  
Thus the plot will in general consist of a superposition of various boxes 
and wedges, hereafter designated as a `wedgebox' plot. The power of the 
wedgebox plot technique manifests itself in precisely such a situation, 
since, given a sufficient number of events, the endpoints of 
(\ref{edge3body}) and (\ref{edge2body}) may each be cleanly identified, and, 
from the relative densities of easily-defined sectors \cite{cascade}
of the wedgebox plot, production ratios such as
$\sigma(pp \to \widetilde{\chi}_i^0 \widetilde{\chi}_j^0) / 
\sigma(pp \to  \widetilde{\chi}_k^0 \widetilde{\chi}_l^0)$, 
may be inferred (since the expected distribution of individual event 
points within a wedge or box from a particular ${ij}$-combination is fairly
simple to model mathematically \cite{PaigeII}).  This information may then 
be used to constrain the neutralino masses and couplings and hence 
the fundamental MSSM input parameters (MSSM IPs) of the neutralino 
mass matrix. Wedgebox plots are hence superior to more traditional 
one-dimensional invariant mass histograms for this four-lepton signature.

Neutralino decay modes other than those included in 
(\ref{basic-decaysZ}) and (\ref{basic-decaysL}) are possible.
Firstly, a neutralino may not decay to the $\widetilde{\chi}_1^0$ LSP
as shown in these reactions, but rather to an intermediate mass 
neutralino, as in  
$ \widetilde{\chi}_4^0 \to \widetilde{\chi}_3^0 +  \ell^+\ell^-$,
$\widetilde{\chi}_4^0 \to \widetilde{\chi}_2^0 +  \ell^+\ell^-$,
or $\widetilde{\chi}_3^0 \to \widetilde{\chi}_2^0 +  \ell^+\ell^-$.
This additional daughter neutralino (or neutralinos) would subsequently 
decay to the $\widetilde{\chi}_1^0$ without producing any more leptons.
The significant presence of such decay chains would introduce `stripes' 
in the wedgebox plot, further enriching its structure:  including the 
possibility of these stripes leads to 178 distinct\footnote{For example, 
$\widetilde{\chi}_2^0 \widetilde{\chi}_2^0$, 
$\widetilde{\chi}_3^0 \widetilde{\chi}_3^0$
and $ \widetilde{\chi}_4^0 \widetilde{\chi}_4^0$ processes each separately
give a box, so a wedgebox plot containing only 
$\widetilde{\chi}_2^0 \widetilde{\chi}_2^0$ is not `distinct' $\,$ 
from a wedgebox plot containing only one of the other two processes.} 
wedgebox plots within the MSSM framework.  Typically though, such extended 
decay chains are unimportant, or at least sub-dominant.
Four-lepton decays from a single neutralino $\widetilde{\chi}_i^0 \to 
\ell^+\ell^- \widetilde{\chi}_k^0 \to \ell^+\ell^-  \widetilde{\chi}_1^0 
{\ell^{\prime}}^+{\ell^{\prime}}^-$ (here the aforementioned daughter 
neutralino does yield a lepton pair from its decay while the other
production neutralino yields no leptons) are also possible, but their 
rates of occurrence are smaller yet.
With inclusion of these stripes, and in the limit of infinite luminosity,
a wedgebox plot from the LHC would consist of a $6 \times 6$
checkerboard in the $M( e^+e^-)-M( \mu^+\mu^-)$-plane (where the 
location of the lines are related to the six possible mass differences 
between the four MSSM neutralinos).
However the actual integrated luminosity of the LHC is limited, by a 
conservative estimate, to roughly $300\, \hbox{fb}^{-1}$ over its 
lifetime, and generally this will not be enough to resolve the full 
checkerboard.  
Instead a specific combination of boxes and wedges will be observed 
in the wedgebox plot based on the dominant production modes for neutralino 
pairs and the dominant neutralino decay modes.  Identifying these dominant 
modes will strongly constrain the MSSM IPs.

Secondly, and more worrisome from the point of view of the present analysis, 
are processes involving charginos.  
A neutralino may decay to the LSP via an intermediate chargino: 
$\widetilde{\chi}_i^0 \to \ell^+\nu +  \widetilde{\chi}_1^- 
\to \ell^+\nu {\ell^{\prime}}^-{\bar{\nu}}^{\prime} + 
\widetilde{\chi}_1^0$.
In such decays, hereafter designated as `mavericks,'  the dilepton
invariant masses are not simply bounded as in reactions
 (\ref{edge3body}) and (\ref{edge2body}).
Fortunately, such mavericks generally constitute a small minority of the 
events (especially for choices of the MSSM IPs which will be found
to be of particular interest) leading to a diffuse `halo' on a wedgebox
plot which is superimposed on the desired sharp box and wedge structure.
Also, the $e^+e^- \mu^+\mu^- \, + \, \slashchar{E}$ final state may
result from $\widetilde{\chi}_i^{\pm}\widetilde{\chi}_j^{\mp}$
chargino pair production, with
$\widetilde{\chi}_i^{\pm} \to \ell\ell\ell^{\prime} X + 
\widetilde{\chi}_1^0$ 
and
$\widetilde{\chi}_j^{\mp} \to \ell^{\prime} Y + \widetilde{\chi}_i^0$ 
(where $X$ and $Y$ are SM final state particles other than $\ell$s, 
typically including neutrinos, and intermediate states may involve 
charged sleptons or sneutrinos).  Such `3+1' events are also lumped into 
the maverick category.  Thus a maverick event is any $e^+e^- \mu^+\mu^-$
event where members of a same-flavor lepton pair arise dis-jointly rather 
than as in\footnote{For `stripe' events, which are not mavericks,
$\widetilde{\chi}_1^0$ would be
replaced by $\widetilde{\chi}_j^0$, $j \ne 1$. in (\ref{basic-decaysZ}) 
or (\ref{basic-decaysL}).} 
(\ref{basic-decaysZ}) or (\ref{basic-decaysL}).
Chargino-neutralino production may
yield final states with five charged leptons of four charged leptons and
a charged quark pair (typically leading to jets) to balance change.
For the former, if the extra lepton is too soft or not isolated or lost 
down the beam pipe,  or, for the latter, if a jet cut fails to exclude 
the event, then chargino-neutralino production may also yield 
$e^+e^- \mu^+\mu^- \, + \, \slashchar{E}$ events.
Charginos, especially $\widetilde{\chi}_2^{\pm}$, may also decay into 
unstable neutralinos (rather than the other way around as above):
$\widetilde{\chi}_i^+ \to \widetilde{\chi}_j^0 + W^+ \to 
\ell^+\ell^- \widetilde{\chi}_1^0+ q \bar{q}^{\prime}$.  Here
this dilepton invariant mass would fit into the expected framework,
so as not to interfere with endpoint studies  
(though presence of such a process would skew attempts to discern  
neutralino pair production rates from event population studies; note 
also the presence of quarks that may yield unacceptable jet activity,
or, if the $W^+$ decays leptonically, an extra lepton would need to 
be lost as above). Another possibility is 
$\widetilde{\chi}_2^{\pm} \to \widetilde{\nu} \ell^\pm 
\to \widetilde{\chi}_1^{\pm}  \ell^\mp  \ell^\pm 
\to  \widetilde{\chi}_1^{0}  \ell^\mp  \ell^\pm W^\pm$
which would give kinematic
edges similar to (\ref{edge3body}) or (\ref{edge2body}) 
(replacing $m_{\widetilde{\chi}_{1,i}^{0}}$ with 
$m_{\widetilde{\chi}_{1,2}^{\pm}}$ and
$m_{\widetilde{\ell}}$ with  $m_{ \widetilde{\nu}}$).
The presence of the above decay chains in an event 
would not invoke the maverick designation. 
Again, though, such processes are expected to have only 
modest rates in regions of phenomenological interest.

As noted above, colored sparticle masses may be pushed up above or around 
the TeV scale to prevent production rates from the cascade channel
(see Fig.\  \ref{inoinoprod}$(c)$) from swamping the other production 
modes.  In \cite{cascade} it was shown in full event generator level 
simulations that ${\sim}500\,\hbox{GeV}$ squarks and gluinos led to the 
overwhelming domination of the cascade channel for the 
$e^+e^- \mu^+\mu^- \, + \, \slashchar{E} + jets$ signature.
Backgrounds were found to be nominal and signal rates high enough to
produce crisp wedgebox plots over a large range of the MSSM IPs associated 
with neutralino characteristics.
This study also showed that, as expected, jet activity is virtually
always associated with cascade events.  Thus a limit on the maximum
number of jets or on the maximum allowable jet energy in an event can
remove most of the cascade events while leaving many of the direct
and Higgs-mediated events (one may speak of demanding that the events be 
`hadronically quiet').  As seen above, such a cut may also reduce the 
effects from maverick events.

This would leave the direct and Higgs-mediated avenues to disentangle.
Note that for both these avenues the two neutralinos arise from the decay 
of a single fundamental particle, whereas in the cascade avenue the 
neutralinos are produced independently (and possibly from decays of 
different colored sparticles --- {\it e.g.}, one neutralino from a gluino 
and the other from a particular species of squark). 
Thus couplings of the EW sector of the MSSM
(excluding those associated with sleptons for the moment),
which are presumably determinable solely from the EW MSSM inputs to the
neutralino mixing matrix, are better scrutinized via a sample of events
from the direct and Higgs-mediated avenues with the cascade avenue
events filtered out.
Study \cite{ha4l} focused on the Higgs-mediated avenue and found that
direct avenue production formed a background to the sought-for 
heavy Higgs boson signals that was difficult if not impossible to remove by 
any set of kinamatical cuts.  To focus instead on the direct channel, one 
could by hand simply choose the Higgs input parameter (generally 
chosen as the pseudoscalar Higgs mass, $m_A$) large enough (in the vicinity 
of a TeV) to shut down the Higgs-mediated avenue.  Nature may not respect 
this choice though.  The present study avoids these dilemmas simply by 
not attempting to cut away either avenue:  the wedgebox plot consists of a 
superposition of shapes from each of the two different avenues.  Each 
avenue may contribute different shapes, if so signaling their respective 
presences, and, for favorable choices of the neutralino-governing MSSM IPs 
(as will be delineated herein), three kinematic edges 
--- as per (\ref{edge3body}) or (\ref{edge2body}) --- may be seen, 
strongly constraining the neutralino masses and IPs.

The remainder of the paper has the following format:
in Section \ref{sec:paramscan} the MSSM IP space is scanned
for the inclusive rate, that is, the rate before the imposition of any 
kinematical cuts, of the neutralino pair-produced 
$ e^+e^- \mu^+\mu^- \slashchar{E} $ signature via the direct and
Higgs-mediated avenues to identify regions of the space where the 
signature is potentially observable.  Guided by these estimations, Section 
\ref{sec:mc} then follows with more detailed full 
event generator Monte Carlo (MC) simulations to carefully analyze the 
salient regions of the MSSM IP space.  Results from the parameter space 
scans and the MC simulations are further expounded upon in 
Section \ref{sec:disc}, and finally Section \ref{sec:conc} gives 
conclusions to be drawn from this work.

\section{Parameter Space Scans}
\label{sec:paramscan}

Before running a full event generator MC simulation of neutralino pair 
production at selected points in the MSSM IP space, it is efficient to
first obtain some estimates of the typical signal and background rates.
Here signal refers to direct production or  Higgs-mediated avenues
of neutralino pair production, 
$ pp \to Z^* \to \widetilde{\chi}_i^0 \widetilde{\chi}_j^0$ or
$ pp \to H^0,A^0 \to \widetilde{\chi}_i^0 \widetilde{\chi}_j^0$.
If the cascade avenue is shut down either by making the colored sparticles 
very massive or by an appropriate jet cut, then the main MSSM backgrounds
are from chargino and slepton production.  
Processes not studied in this initial analysis 
therefore include minor players such as 
$t\overline{t}h$, $tH^\pm$, $tbH^\pm$, {\it etc.}, 
and all SM backgrounds.
Though these all may lead to a
$e^+e^- \mu^+\mu^- \, + \, \slashchar{E}$ signature,
it was shown in  MC studies\cite{ha4l} that they contribute
negligibly after a suitable set of cuts (namely the ones we will
employ in this work): in particular
 SM processes may be virtually
eliminated by demanding a sufficient amount of missing energy, hard 
leptons that are isolated, and limits on jet activity --- save for 
$Z^0Z^{0*}$-induced events, which lead to a few dozen events along the
`Z-lines' on a wedgebox plot; so at worst these lead to
Z-line enhancement.

The MSSM IPs that factor directly in the neutralino and chargino mixing 
matrices are
$\tan\beta$, the ratio of the Higgs boson vacuum expectation values, 
$\mu$, the SUSY higgsino mass parameter, 
and $M_2$, the soft SUSY-breaking 
$SU(2)_{\hbox{\smash{\lower 0.25ex \hbox{${\scriptstyle L}$}}}}$
gaugino mass (in what follows, $M_1$, the soft SUSY-breaking
$U(1)_{\hbox{\smash{\lower 0.25ex \hbox{${\scriptstyle Y}$}}}}$
gaugino mass, is assumed to be fixed by $M_2$ and the gauge 
unification constraint $M_1 = 5/3 \tan^2 \theta_W M_2$).
These MSSM IPs are allowed to take values in the ranges:
$2 < \tan\beta < 50$ (upper limit guided by perturbativity),
and  $100\,\hbox{GeV} < \mu,M_2 < 500\, \hbox{GeV}$ --- 
here the lower bound is set to avoid LEP-excluded light
colorless sparticles and the upper bound to avoid heavier neutralinos
( $\widetilde{\chi}_3^0$ and $\widetilde{\chi}_4^0$ ) too 
massive to generate significant production rates.

Higgs-mediated events are sensitive to the pseudoscalar Higgs mass $m_A$: 
as $m_A$ increases, phase space opens up for more 
$\widetilde{\chi}_i^0 \widetilde{\chi}_j^0$ 
decay channels; however, the  
cross section for $pp \to H^0,A^0$ drops precipitously.
Thus the preferred range for a potentially meaningful contribution
from the Higgs-mediated avenue is 
$300\, \hbox{GeV} \; \lsim \; m_A \; \lsim \; 700\, \hbox{GeV}$.

Also of crucial importance to the neutralino decays into charged leptons 
are the slepton sector MSSM IPs.  Each flavor generation has two soft 
slepton mass inputs ${m_{\widetilde{\ell}_{L,R}}}_i$ 
($i=e,\mu,\tau$)\footnote{There are also trilinear soft inputs 
$A_{{\ell}i}$, but these always come attached to a Yukawa coupling 
and thus are irrelevant for the first two generations.}  Most models
of SUSY breaking generate little splitting between the inputs of the 
first two generations, and thus, for simplicity, these inputs
are set degenerate.  This assumption makes the wedgebox plots virtually 
symmetric under the interchange of the axes, while relaxing this 
assumption may make the wedgebox plot asymmetric (for instance, the 
`boxes' could become `rectangles').  The third generation stau inputs 
are however distinctive in many SUSY-breaking scenarios, and herein these
inputs are elevated (by hand) by $100\, \hbox{GeV}$ over the
degenerate selectron and smuon mass inputs.  The lighter selectrons and 
smuons then favor, via reaction (\ref{basic-decaysL}), events of the 
signature type over those containing tau leptons.  This enhances the 
signal rates while at the same time reducing one additional source of 
maverick events (stemming from events with leptonic tau decays). 
Conversely, measuring the asymmetry and maverick halo density of the 
observed wedgebox plot would provide information about the slepton sector  
mass inputs. 
This leaves two parameters from the slepton sector to vary:  the 
degenerate soft SUSY-breaking mass input for the right sleptons of the 
first two generations (the superpartners of the right-handed electron 
and muon), $m_{\widetilde{\ell}_{R}}$, and the corresponding mass input
for the left sleptons, $m_{\widetilde{\ell}_{L}}$.  If both these 
slepton masses are set very high, then neutralino decays via gauge 
bosons as in reactions (\ref{basic-decaysZ}) totally dominate 
and the leptonic branching ratios (BRs) of the neutralinos are
simply those of the SM gauge bosons.  This is insufficient for generating
enough events for detection of the signature.  Thus positive results in 
this work depend on sleptons being reasonably light 
( $\lsim \; 350\, \hbox{GeV}$ ) --- a condition that fits  
comfortably with the neutralino MSSM IPs under consideration.
These light sleptons will then enhance the leptonic BRs of the 
neutralinos \cite{BaerTata}.
In addition to generating reactions like (\ref{basic-decaysL}),
light slepton mass inputs can also generate   
\begin{eqnarray} 
\widetilde{\chi}_i^0
\to \bar{\nu} + \tilde{\nu}
\to \bar{\nu}\nu + \widetilde{\chi}_1^0 
\label{spoilerdecay}
\end{eqnarray}
decay chains which act as spoiler modes to kill the $4\ell$ signal.
If $m_{\tilde{\nu}} < m_{\widetilde{\chi}_2^0} < 
m_{\widetilde{\ell}^{\pm}}$,
then $\widetilde{\chi}_2^0$ mainly decays via an on-shell sneutrino
and its BR into a pair of charged leptons is highly suppressed.
If SUSY-breaking processes respect 
$SU(2)_{\hbox{\smash{\lower 0.25ex \hbox{${\scriptstyle L}$}}}}$
symmetry, then the sneutrino and left charged lepton of a given flavor 
have the same soft input mass parameter; however, D-terms break this
degeneracy to a limited extent.   Since the sneutrino mass is thus tied 
to $m_{\widetilde{\ell}_{L}}$, lowering  $m_{\widetilde{\ell}_{R}}$ 
relative to $m_{\widetilde{\ell}_{L}}$ tends to improve the signal rate.

A private code was used normalized by cross-sections input from the 
event generator ISAJET \cite{ISAJET} to perform a scan over the 
$\mu$ and $M_2$ neutralino IPs for the signature 
$\sigma(pp \to  X) \times B.R.( X \to  e^+e^-   \mu^+\mu^-)$
where $X$ represents the intermediate states (as in Fig.\ \ref{inoinoprod} 
$(a)$ or $(b)$ in the case of the signal).
Other MSSM IPs were fixed as follows:  $\tan\beta = 10$, 
$m_A = 600\, \hbox{GeV}$, $m_{\tilde{g},\tilde{q}} = 1000\, \hbox{GeV}$,
$m_{\tilde{e},\tilde{\mu} } = 150\, \hbox{GeV}$,
$m_{\tilde{\tau}} = 250\, \hbox{GeV}$, and vanishing soft $A$-terms.
In this initial parton-level analysis, the mere presence of exactly the 
four leptons in the signature is all that is required with no demands 
whatsoever made upon their kinematical properties ({\it e.g.}, transverse 
momenta or pseudorapidity).  Any effects from the underlying spectator 
event are neglected.  By contrast, in the full event generator MC analysis 
to follow appropriate cuts on the leptons' kinematical properties will
be applied, meaning that the numerical results of the initial analysis
are over-estimates.  The initial analysis also demands no quarks in the 
final state, where only particles resulting from the primary 
parton-level interaction are taken into account.  On the other hand,
in the full event generator MC analysis at the very least quark remnants 
from the colliding protons must yield quarks in the final state (though 
these typically lie close to the beam axis).  Thus
at best a lower bound can be set upon hadronic or jet activity in the 
final state, which would tend to make the results of the initial analysis
under-estimates of the event generator MC results.  Of these two 
differences between the two analyses, the former effect is expected to be 
more significant.  Thus the results from the initial analysis may be 
treated as upper bounds of what may be expected from the more thorough
event generator MC studies.

\begin{figure}[htb!]
\vskip -1.0cm
\dofig{6.00in}{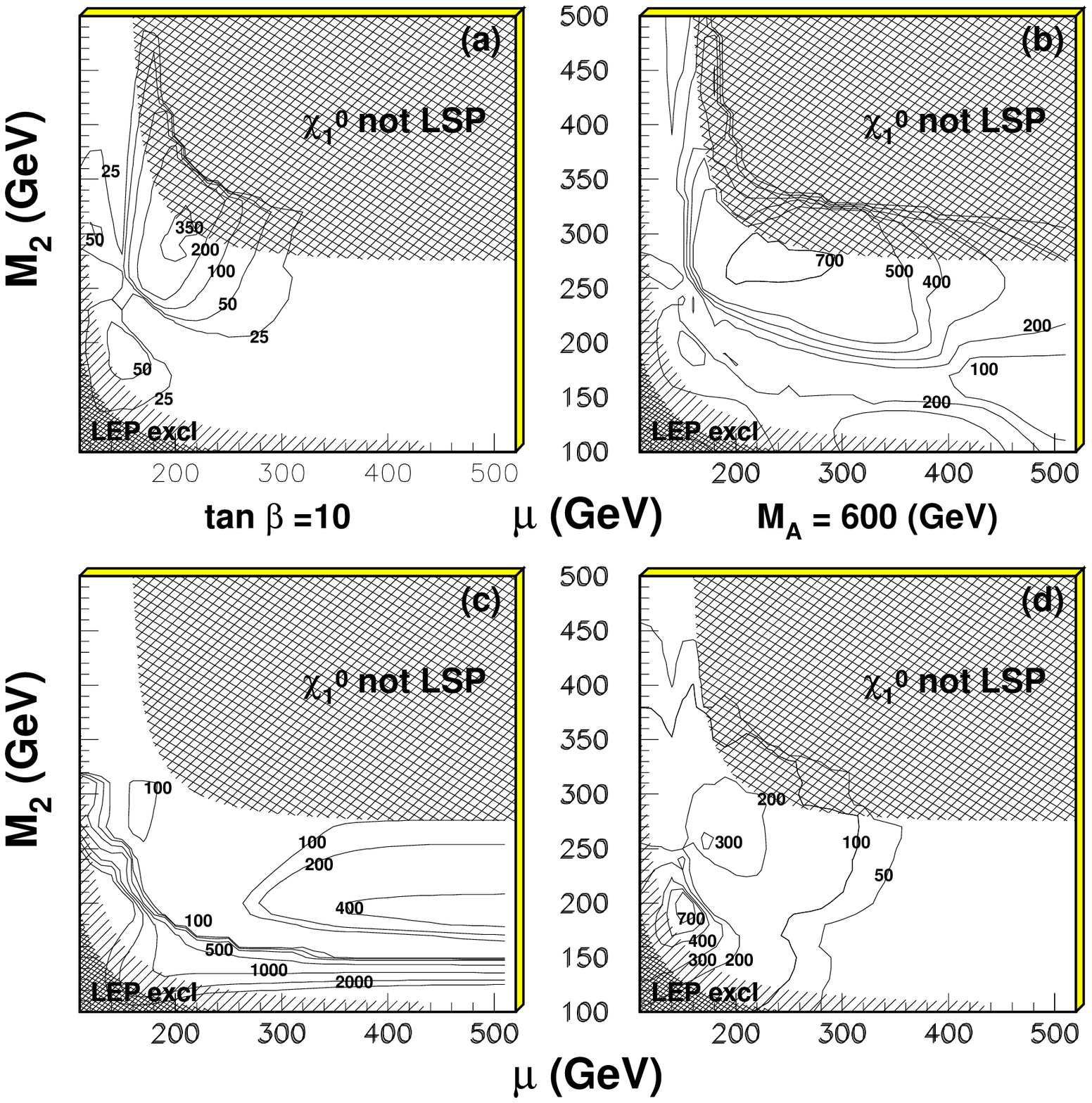}
\vskip -0.4cm
\caption{Number of $ e^+e^- \mu^+\mu^-$ events 
(inclusive rates with no cuts) expected per $100\, \hbox{fb}^{-1}$ of 
integrated luminosity from
{\it (a)} $pp \to \widetilde{\chi}_2^0 \widetilde{\chi}_3^0 $;
{\it (b)} $pp \to H^0 / A^0$;
\newline
{\it (c)} $pp \to  \widetilde{\chi}_i^0 \widetilde{\chi}_j^{\pm} $ and
{\it (d)} $pp \to  \widetilde{\chi}_i^{\pm} \widetilde{\chi}_j^{\mp} $.
Other MSSM inputs are fixed as $\tan\beta = 10$,
$m_A = 600\, \hbox{GeV}$, $m_{\tilde{e},\tilde{\mu} } = 150\, \hbox{GeV}$
and $m_{\tilde{\tau}} = 250\, \hbox{GeV}$.  The uncertainty shown in the 
extent of the LEP excluded region stems from the presence of a relatively
light sneutrino (as discussed in a footnote).
 \label{paramspace1} }
\end{figure}

Fig.\ \ref{paramspace1} shows the results\footnote{Note that in this 
plot, as well as other to follow $\mu > 0$ is chosen.  While analogous
plots for $\mu < 0$ are not quite symmetric to the $ \mu > 0$ plots
shown here, substantive differences are few with the same features 
appearing at slightly shifted values of $ | \mu | $.}, 
assuming an integrated
luminosity of $100\, \hbox{fb}^{-1}$.  The lower and upper shaded areas 
are excluded by LEP searches (restricting\footnote{For physical sneutrino 
masses
below $200\, \hbox{GeV}$, destructive interference from a $t$-channel
sneutrino exchange diagram with the normal $s$-channel diagram for  
$e^+e^-$-collider chargino pair production lowers the bound given by
LEP experimental groups \cite{PDB2006} from
$m_{  \widetilde{\chi}_1^\pm} > 103\, \hbox{GeV}$ (singly hatched bound
on plots) to
$m_{  \widetilde{\chi}_1^\pm} > 85\, \hbox{GeV}$ (doubly hatched bound
on plots).  A true experimentalist's bound for the MSSM IP sets considered
herein would thus be expected to lie somewhere within the singly hatched
zone.} $m_{  \widetilde{\chi}_1^\pm}$) 
and cosmological/dark matter considerations ({\it i.e.}, require 
$\widetilde{\chi}_1^0$ to be the LSP), respectively.  
Plot $(a)$ shows what may be expected from the direct channel.  
Of the six possible  $\widetilde{\chi}_i^0 \widetilde{\chi}_j^0$
pairs ($i,j =2,3,4$), 
only the $ \widetilde{\chi}_2^0 \widetilde{\chi}_3^0 $ combination 
leads to a significant number of events (set as 100 events). 
Phase space suppression renders the 
$ \widetilde{\chi}_i^0 \widetilde{\chi}_4^0$ channels negligible.
The rate for $ \widetilde{\chi}_2^0 \widetilde{\chi}_2^0$ is suppressed 
since, in the pertinent region of the $\mu , M_2$ parameter space, the
$ \widetilde{\chi}_2^0$ has approximately equal higgsino components and 
the $ Z\widetilde{\chi}_2^0 \widetilde{\chi}_2^0$
coupling\footnote{In the notation of \cite{HaberKane}, this term is
$<Z|\widetilde{\chi}_i^{0}\widetilde{\chi}_j^0>
=\left(g/2cos\theta \right) Re(N_{i3}N_{j3}^{*}-N_{i4}N_{j4}^{*}) \;\; .$
The crucial minus sign in this equation arises from the different 
hypercharges of the two MSSM Higgs doublets.  If $i=j$, this leads to a
strong tendency for the two terms to cancel each other.  However, for 
$i=2$ and $j=3$, as in direct $\widetilde{\chi}_2^0 \widetilde{\chi}_3^0$
production, the signs of either $N_{23}$ and $N_{33}$ {\em or}
$N_{24}$ and $N_{34}$ --- {\em but not both} --- are opposite over much 
of the interesting region of the MSSM IP space, and enhancement rather 
than cancellation ensues.}  vanishes due to the cancellation between the 
contributions from these two higgsino components.
An analogous suppression occurs with
the $\widetilde{\chi}_3^0 \widetilde{\chi}_3^0$ mode, along with
substantial phase space suppression.
Note that there are too few events at either high $M_2$ and/or high 
$\mu$ to meet the significance criterion due to the small size of
$\sigma(pp \to \widetilde{\chi}_2^0 \widetilde{\chi}_3^0)$. 
In the remaining portion of the plane sneutrino spoiler modes
--- which confound  
$\widetilde{\chi}_2^0 \to \ell^+ \ell^- \widetilde{\chi}_1^0$
decays --- carve out a valley in the event-rate topology, seen in 
the $(\mu , M_2)$ MSSM IP plane as a
more or less hyperbolic swath passing through 
$(\mu , M_2) = (200\, \hbox{GeV} , 200\, \hbox{GeV})$,
leaving two peaks at 
$(\mu , M_2 ) \simeq (200\, \hbox{GeV} , 300\, \hbox{GeV})$
and 
$(150\, \hbox{GeV} , 175\, \hbox{GeV})$\footnote{Though not very 
discernible on the plots, there is also a very narrow bridge of 
high rates centered on 
$(\mu , M_2 ) \simeq (155\, \hbox{GeV} , 245\, \hbox{GeV})$.
This is where the sneutrino coupling to $\widetilde{\chi}_2^0$
dies, turning off the most important spoiler mode.}.

Plot $(b)$ gives the results for the Higgs-mediated channels. Rates 
everywhere exceed those of the direct  
$\widetilde{\chi}_2^0 \widetilde{\chi}_3^0$ production mode since the 
mechanism suppressing 
$\widetilde{\chi}_2^0 \widetilde{\chi}_2^0$ in the direct channel does not
apply to the Higgs' couplings. 
However the optimal point in the plane is in roughly the same location,
around $(\mu , M_2 ) \simeq (200\, \hbox{GeV} , 275\, \hbox{GeV})$.
This is strongly influenced by the aforementioned sneutrino spoiler 
mechanism choking off the event rate as one moves off this peak.
These features directly follow from 
the choices made for $m_A$ and $\tan\beta$ ($600\, \hbox{GeV}$ and $10$, 
respectively).
As found in \cite{ha4l}, Higgs decays 
to $\widetilde{\chi}_2^0 \widetilde{\chi}_2^0$ tend to dominate 
for larger values of $\mu$.  This means that Higgs-mediated
processes can lead to a box, whereas the direct avenue is 
expected to produce a wedge.  On the other hand, \cite{ha4l} also found 
that decays including the heavier neutralinos $\widetilde{\chi}_3^0$ and 
$\widetilde{\chi}_4^0$ may be very significant or even dominate for
smaller values of $\mu$ (assuming $m_A$ is sufficiently large). 
Thus more complicated wedgebox plots may be expected from  Higgs-mediated
processes at lower values of $\mu$ (and higher values of $m_A$).

The most significant SUSY backgrounds involve charginos.
Plots $(c)$ and $(d)$ of Fig.\ \ref{paramspace1} display expected 
$5-$ and $4$-lepton event rates from 
$pp \to  \widetilde{\chi}_i^0 \widetilde{\chi}_j^\pm$
and 
$pp \to  \widetilde{\chi}_i^{\pm} \widetilde{\chi}_j^{\mp}$,
respectively.  Here the $\widetilde{\chi}_i^0 \widetilde{\chi}_j^\pm$
pair is required to produce five leptons, and then one lepton would
have to be `lost.'  Losing the extra lepton is {\em not} taken into 
account in the rates shown in $(c)$, and thus rates shown for this 
process are certainly considerably over-estimated.
Nonetheless, the plot clearly shows that the largest rates from
$\widetilde{\chi}_i^0 \widetilde{\chi}_j^\pm$ should come at 
low values of $M_2$ (with some preference also for higher values of 
$\mu$).  This is not a region where the direct and 
Higgs-mediated neutralino-pair production processes are expected to 
yield enough events to sufficiently populate a wedgebox plot.  Thus 
at worst $\widetilde{\chi}_i^0 \widetilde{\chi}_j^\pm$ processes  
would contribute a small minority of the events in a neutralino-pair-induced
wedgebox
plot.  As seen from plot $(d)$, chargino pair production is expected to
generate a fair number of four lepton events (typically mavericks) 
which might act to cloud the neutralino-based features of the wedgebox 
plot.  Light charginos are
generally expected to have larger cross-sections at the LHC than 
neutralinos.  Fortunately,  
$ \widetilde{\chi}_1^+ \widetilde{\chi}_1^- $-production
can almost never generate the four-lepton final state.  Requiring
processes involving the heavier chargino pushes the location for 
optimal rates from chargino-pair production to quite low values of 
$M_2$ and $\mu$.  This is mostly non-overlapping with the sweet spot 
for neutralino-pair-production processes; however, a secondary maxima
in the chargino-pair rates is seen at   
$(\mu , M_2 ) \simeq (200\, \hbox{GeV} , 250\, \hbox{GeV})$,
and this is in the region where a neutralino-pair-induced wedgebox 
plot would be viable.   One possible method for alleviating this
problem (not implemented in this work) would be to examine  
$\ell^+\ell^- \ell^+ \ell^{\prime -}$ events 
(since with $\widetilde{\chi}_i^+\widetilde{\chi}_j^-$
one chargino is expected to produce three leptons while the other
chargino produces only one, while with 
$\widetilde{\chi}_i^0\widetilde{\chi}_j^0$
each neutralino should, with rare exceptions, produce a pair of 
same-flavor leptons),
which should have rates equivalent to $\ell^+\ell^- \ell^+ \ell^-$,
as the basis for a chargino-pair event subtraction 
scheme~\cite{EEEEMUMUMUMU}.  

Finally, slepton production also comprises a potentially large background.
In \cite{ha4l}, four-lepton signature events from slepton-pair production
were   found to be even harder to cut away from the desired Higgs-mediated
signal than events from direct avenue neutralino pair production, 
though in that case only enough Higgs-mediated signal events were
sought to claim a signal of $\sim 20$ events after all
cuts. In this work, on the other hand, 
hundreds of events are needed. We discuss
this issue at more length in the following section, where it is found
that sleptons contribute significantly only at very low values of $\mu$
or $M_2$; {\it i.e.}, in regions where signal rates are small.
However, sleptons are always of paramount importance as intermediates in the
decays of the neutralinos to the desired charged leptons.

\section{Monte Carlo Event Generator Analysis}
\label{sec:mc}

The  HERWIG 6.5 \cite{HERWIG65} MC package (which obtains its
MSSM input information from ISASUSY \cite{ISAJET} through the ISAWIG  
\cite{ISAWIG} and HDECAY \cite{HDECAY} interfaces) is employed to 
generate realistic LHC events.  The CTEQ 6M \cite{CTEQ6} set of parton
distribution functions is used with top and bottom quark masses
set to $m_t=175\, \hbox{GeV}$ and $m_b=4.25\, \hbox{GeV}$, respectively.
This is coupled with private programs simulating a typical 
LHC detector environment (these codes have been checked against results in 
the literature).
Assuming an integrated luminosity of $100\, \hbox{fb}^{-1}$,
roughly equivalent to the first few years' data at
the LHC,
the appropriate numbers of events (normalization was according to 
HERWIG-delivered cross-sections) for signal and background processes
were generated at an array of points spanning the 
$(\mu , M_2 )$ plane.

Signature $e^+e^- \mu^+\mu^- $ events are selected according to the 
following criteria:
the event must have exactly four hard ($p_T^\ell > 10, 8\, \hbox{GeV}$
for $e^{\pm},\mu^{\pm}$, respectively), isolated (no tracks of other
charged particles in a $r = 0.3\, \hbox{radians}$ cone around the lepton,
and less than $3\, \hbox{GeV}$ of energy deposited into the  
electromagnetic calorimeter for $0.05\, \hbox{radians} < r < 0.3\, 
\hbox{radians}$ around the lepton) leptons, consisting of exactly
one $e^+e^-$ pair and one $\mu^+\mu^-$ pair.

After identifying signature events, the following cuts are then applied:
\newline
$\bullet$
Substantial missing transverse energy must be present, with
\newline
\phantom{aaaaa} $20\, \hbox{GeV} < \slashchar{E_{T}} < 130\, 
\hbox{GeV}\;$, \phantom{aa} and
\newline
$\bullet$
No jet reconstructed with an energy $E_{jet}$ greater than 
$50\, \hbox{GeV}$.  
\newline
\phantom{aaaaa}
Jets are defined by a cone algorithm with 
$r=0.4$ and must have $|\eta^j|<2.4\,$ . 
\newline
These cuts are sufficient to eliminate all of the SM backgrounds except 
that from $Z^0Z^{0*}$ production.  Gluino and/or squark pair production
events are also removed, leaving only residual SUSY
backgrounds from processes involving charginos 
($p p \to \widetilde{\chi}_i^\pm \widetilde{\chi}_j^\pm \, \hbox{or} \, 
\widetilde{\chi}_i^0 \widetilde{\chi}_j^\pm$ via SM
gauge bosons or Higgs bosons), from charged slepton pair production
($ p p \to \widetilde{\ell}^\pm \widetilde{\ell}^\mp$), and
(making a minor contribution) 
from $pp \to tH^-,\bar{t}H^+$. 

\begin{figure}[t!]
\vskip -0.4cm
\dofig{6.00in}{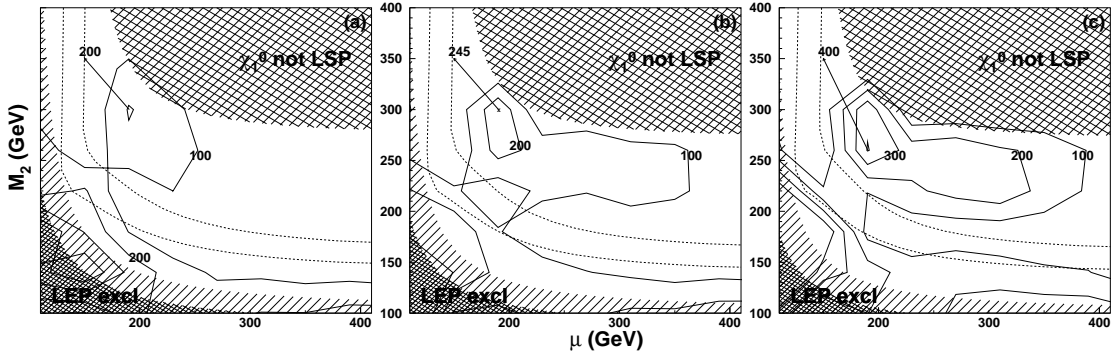}
\caption{Number of $e^+e^-\mu^+\mu_-$ events 
(event generator simulated, after cuts)
at $\tan\beta = 5$ (left),
$\tan\beta = 10$ (middle),
and at $\tan\beta = 20$ (right); $m_A = 600\, \hbox{GeV}$,
$m_{\tilde{\ell}_L} = m_{\tilde{\ell}_R} = 150\, \hbox{GeV}$
($\ell = e, \mu$) and $m_{\tilde{\tau}} =  250\, \hbox{GeV}$.
Assuming an integrated luminosity of $100\, \hbox{fb}^{-1}$.
Region between the two hyperbolic dashed curves is where the 
sneutrino spoiler modes cut heavily into event rates.
\newline
 \label{MCvartanB} }
\end{figure}
\begin{figure}[h!]
\vskip -0.7cm
\dofig{4.00in}{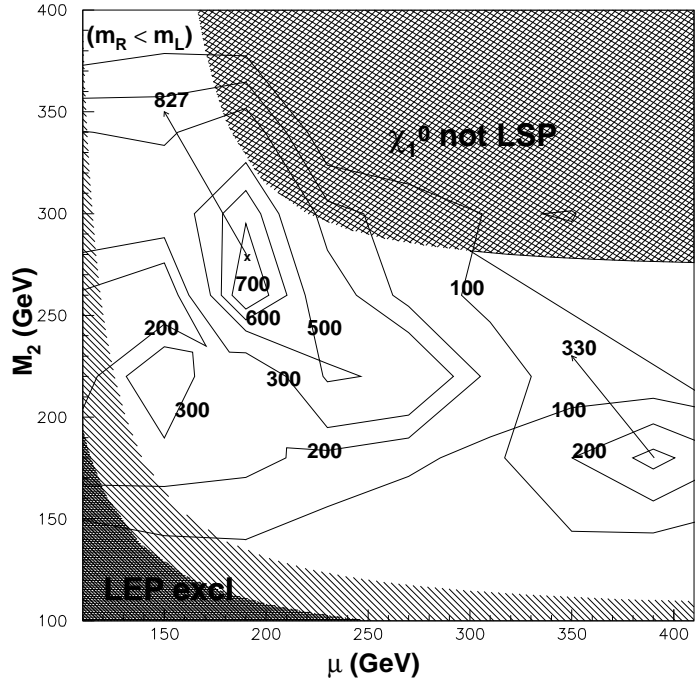}
\vskip -0.3cm
\caption{Number of $e^+e^-\mu^+\mu_-$ events 
(event generator simulated, after cuts)
with $\tan\beta = 10$,
$m_{\tilde{\ell}_L} = 200\, \hbox{GeV}$ and 
$m_{\tilde{\ell}_R} = 150\, \hbox{GeV}$
($\ell = e, \mu$) and $m_{\tilde{\tau}} =  250\, \hbox{GeV}$; 
$m_A = 600\, \hbox{GeV}$.
Assuming an integrated luminosity of $100\, \hbox{fb}^{-1}$.
 \label{MCslepvar} }
\vskip -0.3cm
\end{figure}

Fig.\ \ref{MCvartanB} shows contour plots in the $(\mu, M_2)$-plane
of the number of events, due to the combined signal and background 
processes given in Fig.\ \ref{paramspace1} (plus the slepton background),
 expected to pass the above set 
of cuts, assuming an integrated luminosity of $100\, \hbox{fb}^{-1}$.  
Two  `islands' where the number of expected events swell to over $100$
appear.
As expected, rates are down (roughly by factors of from two to 
four) from the bounding estimates given Fig.\ \ref{paramspace1}, but
the general features match well those expected based on the parameter
space scan.  The location of the upper island corresponds fairly well with
the locations of the maxima from direct and Higgs-mediated neutralino
pair production in  Fig.\ \ref{paramspace1}.  Also according to 
Fig.\ \ref{paramspace1}, the lower island, which is partially covered by 
the LEP excluded region, is situated where both the direct and 
Higgs-mediated signal processes {\em and} the chargino-related background 
processes produce substantial numbers of events.  
The numbers in  Fig.\ \ref{paramspace1} would seem to 
indicate that the lower island should be totally dominated by 
chargino-related events; however, recall that plot $(c)$ in that
figure does not take into account the fraction of the time the extra
fifth lepton from $pp \to  \widetilde{\chi}_i^0 \widetilde{\chi}_j^{\pm}$
is lost.  After cuts, signal and background rates on this lower island
are found to be comparable, though chargino-related plus slepton 
background events
still typically contribute a majority of the events. In particular, in a
thin strip of points hugging the $\mu$ axis slepton pair production,
which yields four leptons via a `3+1' process,
\begin{eqnarray}
 \tilde{\ell}^\pm  \to \ell'^\pm \widetilde{\chi}_2^0 
   \to   \ell'^\pm  \ell^\mp  \ell^\pm \widetilde{\chi}_1^0 \\ \nonumber
 \tilde{\ell}^\mp \to  \ell'^\mp \widetilde{\chi}_1^0 \,\,\,\, ,
\label{3+1}
\end{eqnarray}
yields almost all the events since direct and Higgs-mediated modes 
are cut back by increasingly heavy neutralino masses.
A corresponding strip also extends along the $M_2$ axis, but here sleptons 
prefer to decay to $\widetilde{\chi}_3^0$ until it becomes kinematically
more favorable to decay to charginos 
($m_{\widetilde{\chi}_3^0} > m_{\tilde{\ell}^\pm} > m_{\widetilde{\chi}_1^\pm}$)
which do not yield the desired 4-lepton final state; 
therefore for the particular slepton masses employed in these scans this 
strip terminates near  $M_2 \sim 300\, \hbox{GeV}$.
The plots in Fig.\ \ref{MCvartanB} also illustrate that rates
around the upper island in the $(\mu , M_2 )$ plane rise with $\tan\beta$; 
mostly this is due to an accompanying increase in the Higgs bosons'
production rates -- Higgs-mediated events rise from less than $10\%$ 
of the total at $\tan\beta=5$ to $60\%$ at  $\tan\beta=20$ --- but 
$\tan\beta$-induced changes in the neutralino \& chargino masses and 
couplings also play a role.  If the SUSY-breaking stau mass 
inputs are set equal to or slightly above the inputs of the first two 
slepton generations, then rates of all signal processes will fall 
precipitously after some high $\tan\beta$ limit is reached.  This is due 
to mixing in the stau sector driving down one of the physical stau masses 
leading to sparticle and heavy Higgs boson decays predominantly filled 
with tau-leptons.
The complex interplay of $\tan\beta$ with the observed masses and 
couplings underscores the difficulty in going from an observed number 
of events to predictions for MSSM IP values.
Note however that the gross appearance and location of the 
maxima/`islands' where a sufficient number of events are produced is 
little altered.

Note also the continued appearance of the hyperbolic swath in the 
$M_2$ {\it vs.} $\mu$ plots.
As in  Fig.\ \ref{paramspace1}, this is
caused by sneutrino spoiler modes which are expected to dominate
in the region between the dotted lines in  Fig.\ \ref{MCvartanB}.  
The strength of said spoiler modes varies with the input slepton mass 
parameters chosen.
These include (see \cite{cascade}): $\tan\beta$, trilinear soft $A$-terms 
whose effects are insignificant for the first two generations, and the 
soft slepton mass inputs.  Results with the canonical choice of
$m_{\tilde{\ell}_L} = m_{\tilde{\ell}_R} = 150\, \hbox{GeV}$
(used in all of the other figures presented in this work)
seen in the center ($\tan\beta=10$) plot of  Fig.\ \ref{MCvartanB} may be 
compared to those from elevating $m_{\tilde{\ell}_L}$ to $200\, \hbox{GeV}$ 
in  Fig.\ \ref{MCslepvar}.  
Elevating $m_{\tilde{\ell}_L}$ raises the physical sneutrino masses 
(as well as the masses of the left charged sleptons)
above those of the right charged sleptons, choking off the spoiler modes.
Comparing the two results, we see that the rates when $m_{\tilde{\ell}_L}$
is made heavier actually increase, despite the diminution of the left
charged slepton channel, and the swath of low event rates that was cutting
across the plot has now largely vanished, allowing a third maximum to
surface near  $( M_2 , \mu ) = ( 390\, \hbox{GeV} , 180\, \hbox{GeV} )$.
This third peak is almost entirely due to Higgs-mediated 
$H^0/A^0 \to \widetilde{\chi}_{2}^0 \widetilde{\chi}_{2,3}^0$
modes (direct modes' cross-sections
are too small here): as one moves off this peak to lower $M_2$ or $\mu$, 
intermediate sleptons become off-shell and suppress
$\widetilde{\chi}_2^0 \to \ell^+ \ell^-  \widetilde{\chi}_1^0$ decays, 
 whereas moving
to higher values of these parameters raises the masses of
 $\widetilde{\chi}_{2,3}^0$ and thus
 suppresses 
$H^0/A^0 \rightarrow \widetilde{\chi}_2^0 \widetilde{\chi}_{2,3}^0$ modes.

\begin{figure}[t]
\vskip -1.0cm
\dofig{7.00in}{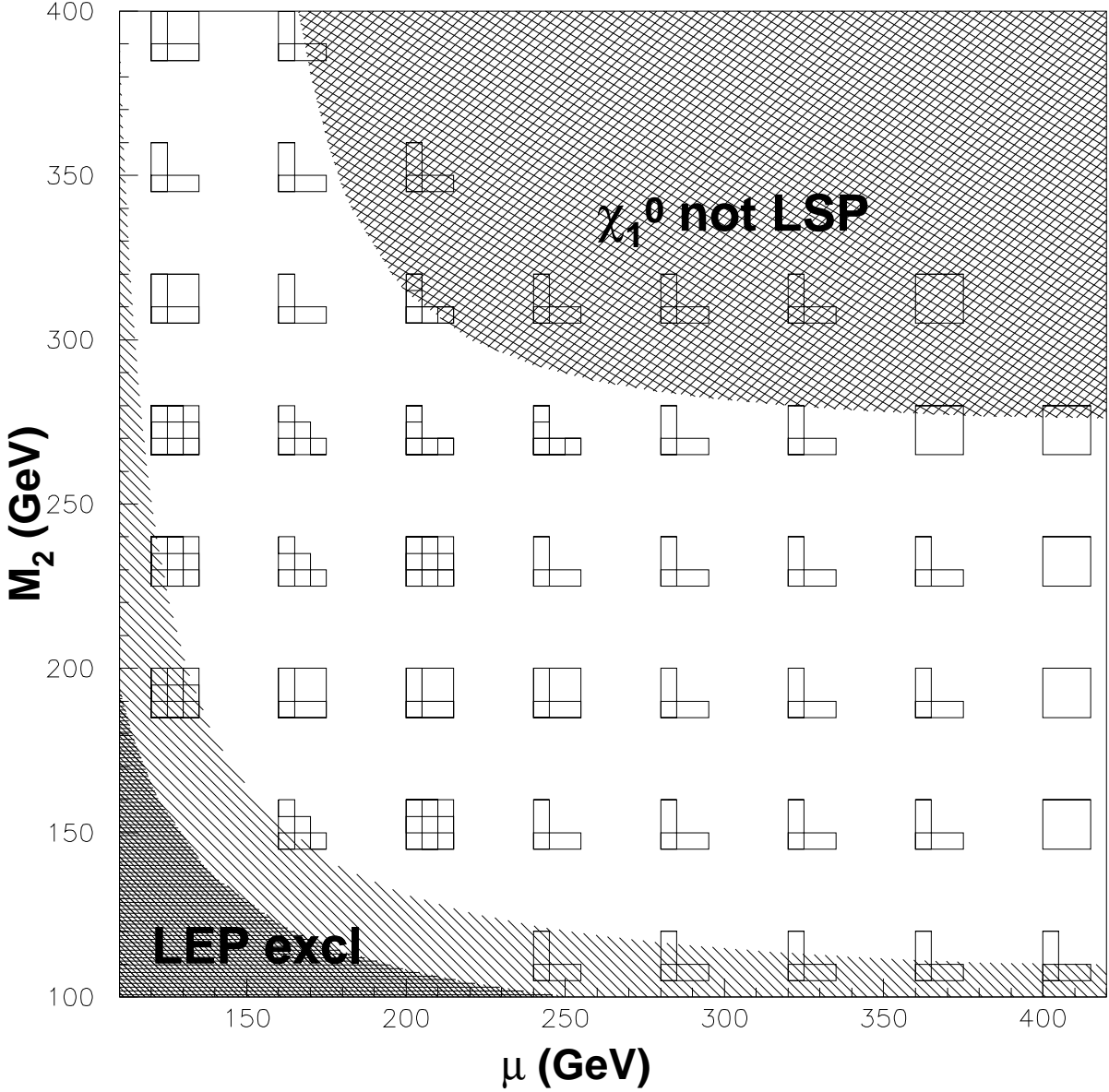}
\vskip -0.1cm
\caption{Wedgebox `map' for $\tan\beta = 20$ and assuming an 
integrated luminosity of $300\, \hbox{fb}^{-1}$.  Idealized
wedgebox plot patterns abstracted from visual inspections of wedgebox 
plots obtained from simulation runs at an array of points spanning the 
parameter space.  Values of other fixed parameters:
$m_A = 600\, \hbox{GeV}$, $m_{\tilde{g},\tilde{q}} = 1000\, \hbox{GeV}$,
and ${m_{\widetilde{\ell}_{L,R}}}_i = 150\, \hbox{GeV}$ ($i=e,\mu$),
$250\, \hbox{GeV}$ ($i=\tau$).
The uncertainty shown in the extent of the LEP excluded region stems from 
the presence of a relatively
light sneutrino (as discussed in an earlier footnote).
\newline
 \label{wedgemap} }
\vskip 0.5cm
\end{figure}

The foregoing demonstrates that regardless of the values of 
$\tan\beta$, $M_A$, and the slepton masses\footnote{Assuming these
are less than $\sim 300\, \hbox{GeV}$, otherwise event rates are too 
low.}, there are always two disjoint regions of high (over $100$ events 
per $100\, \hbox{fb}^{-1}$) rates in the $( M_2 , \mu )$ plane.
Let us now examine what, if any, pattern there is to the wedgebox plots
in these regions.  Taking for definiteness $\tan\beta=20$ (right plot of 
Fig.\ \ref{MCvartanB}), wedgebox plots are generated at an array of 
$(\mu, M_2)$ points, assuming an integrated luminosity of 
$300\, \hbox{fb}^{-1}$, to obtain the `wedgebox map' shown in 
Fig.\ \ref{wedgemap}.
Each symbol in the wedgebox map represents a shape ascertained from visual 
inspection of the corresponding wedgebox plot (explicit examples 
are forthcoming) at that point. Based upon an LHC integrated 
luminosity of $300\, \hbox{fb}^{-1}$, this wedgebox map represents the 
potential of the LHC to correlate neutralino MSSM IPs $\mu$ and $M_2$ with 
an observed wedgebox shape if Nature has chosen $\tan\beta=20$, 
$M_A= 600\, \hbox{GeV}$, $m_{\tilde{g},\tilde{q}} = 1000\, \hbox{GeV}$,
and ${m_{\widetilde{\ell}_{L,R}}}_i = 150\, \hbox{GeV}$ ($i=e,\mu$),
$250\, \hbox{GeV}$ ($i=\tau$).  Fig.\ \ref{wedgemap} is thus a
representative example of a class of $(\mu , M_2)$ wedgebox plots 
that can be made by varying these additional inputs.

\begin{figure}[t]
\vskip -2.0cm
\dofig{4.20in}{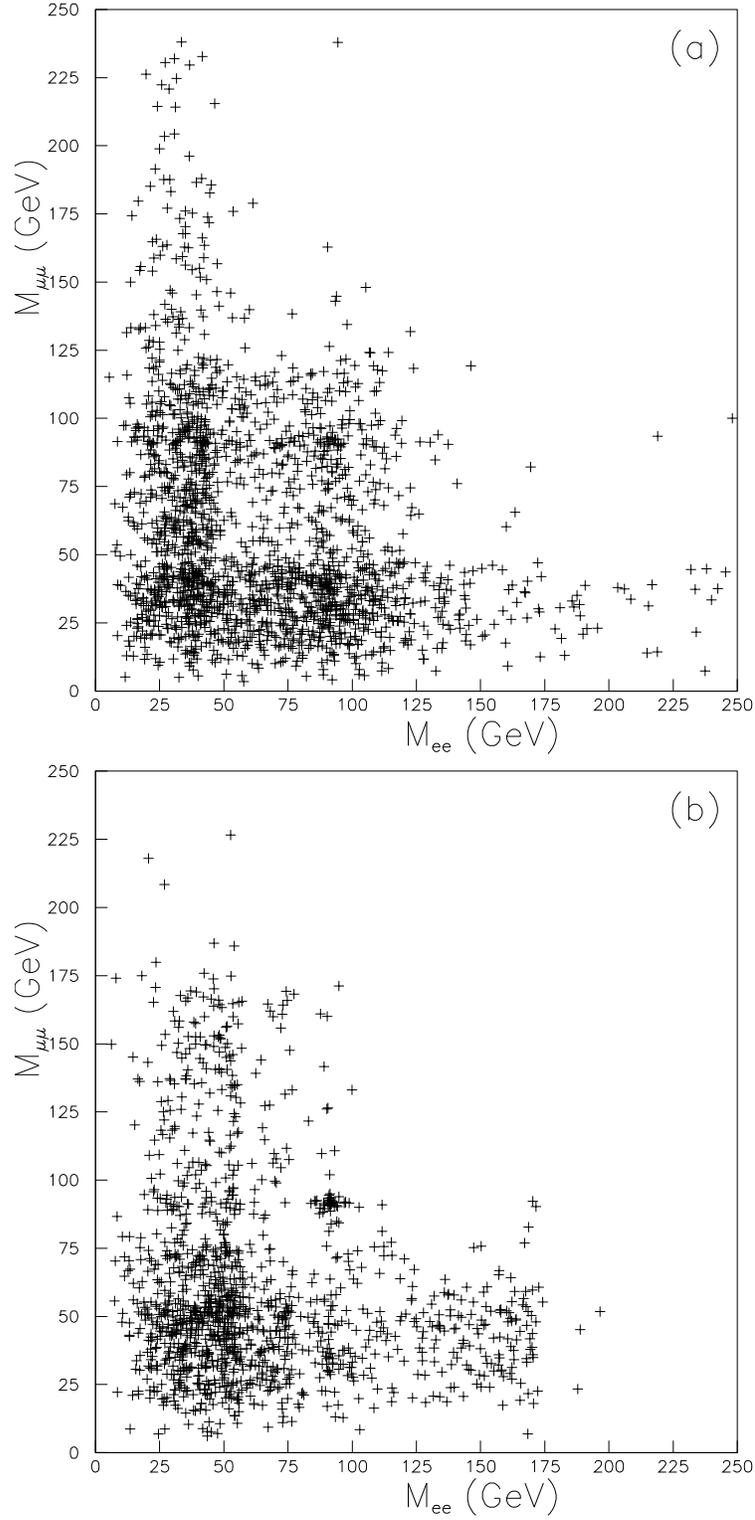}
\caption{Wedgebox patterns for an integrated luminosity of 
$300\, \hbox{fb}^{-1}$ at 
{\it (a)} $(\mu , M_2) = (150\, \hbox{GeV} , 160\, \hbox{GeV})$
and {\it (b)} $(\mu , M_2) = (190\, \hbox{GeV} , 280\, \hbox{GeV})$; 
in both cases $\tan\beta = 20$, $m_A = 600\, \hbox{GeV}$, 
and $m_{\tilde{\ell}_L} = m_{\tilde{\ell}_R} = 150\, \hbox{GeV}$.  
All backgrounds remaining after cuts as described in the text are 
included.  Bin size is $2.5\, \hbox{GeV}$ along each axis.
 \label{Dalitz-exmpls} }
\end{figure}

Toward the lower left-hand corner of Fig.\ \ref{wedgemap}, in the region 
of the lower island, one sees a fairly complicated evolution of shapes, as 
might be expected since since here the direct, Higgs-mediated, 
chargino-related and slepton-pair production modes all 
contribute significantly.  For example, the wedgebox pattern for the 
point $( \mu , M_2 ) = ( 150\, \hbox{GeV} , 160\, \hbox{GeV} )$ is 
depicted in Fig.\ \ref{wedgemap} as a box with a wedge extending out of it.  
The actual wedgebox plot is shown 
in Fig.\ \ref{Dalitz-exmpls}$(a)$.   A sizable fraction of the events are 
chargino-related mavericks ({\it vis \`a vis} Secn.\ 1).
While some kinematical edges are clearly visible, the high fraction of
mavericks makes it especially difficult to connect these with mass
differences in the neutralino spectrum. 
For example, the clustering of points near  $45\, \hbox{GeV}$ is in fact 
a mixture of the
$\widetilde{\chi}_2^0 \to \widetilde{\chi}_1^0$ decay edge through an 
off-shell slepton
($ M_{21}(\ell^+\ell^-)=45.7\, \hbox{GeV}$)
and the $\widetilde{\chi}_3^0 \to \widetilde{\chi}_1^0$ edge through
on-shell sleptons
($32.5\, \hbox{GeV} <  M_{31}(\ell^+\ell^-) 
< 45.5\, \hbox{GeV}$)\footnote{Here it is necessary to make note 
of a small inadequacy in the analysis package used to generate the 
wedgebox plots: a term in the slepton masses from left-right sfermion 
mixing --- the $m^2_\ell \mu^2\tan^2\beta$ term in
Eqn.~(13) of \cite{cascade} --- is neglected in ISASUSY 7.58
$\!\!\!\!\!$ \cite{ISAJET} which feeds the mass values into HERWIG 6.5 
\cite{HERWIG65}.  Neglecting this term, the mass splitting of the
smuons becomes equal to that of the selectrons, and thus is evidently
sometimes under-estimated.  Numerical values given in the text
correctly account for this term, and so may not exactly correspond to what 
is observed in the wedgebox plot figures, though the differences are 
not crucial to the current analysis and discussion.}.
The large width of $M_{31}$ here is due to the proximity of the 
slepton masses to that of $\widetilde{\chi}_3^0$.
Moreover, the edge seen near $115\, \hbox{GeV}$ arises not from a 
neutralino decay, but from 
$\widetilde{\chi}_2^\pm \to \widetilde{\nu} \ell^\pm
 \to \widetilde{\chi}_1^\pm  \ell^\pm  \ell^\mp$ : this decay yields two 
leptons with an invariant mass cutoff of $118.7\, \hbox{GeV}$
determined by the charginos' mass difference. 
Not easily discernible in the plot is a another edge from 
$\widetilde{\chi}_4^0 \to \widetilde{\chi}_1^0$ decays (calculated as 
$141.8\, \hbox{GeV} <  M_{41}(\ell^+\ell^-) < 144.6\, \hbox{GeV}$); 
these mostly arise from Higgs-mediated events.
Finally, the long tail of events along the axes arises from the slepton 
background, since at these small values of $\mu$ and $M_2$ the '3+1' decay 
modes (see (\ref{3+1})) open: a hallmark of the '3+1' modes is a wedge 
with a diffuse tail extending to high invariant masses,
since one lepton pair will have a well-defined invariant mass cutoff 
(usually $m_{\widetilde{\chi}_2^0} - m_{\widetilde{\chi}_1^0}$) while 
the other pair will not. 
The multitude of significant source processes for the event points, the 
quick evolution of wedgebox patterns as one moves around the 
$( \mu , M_2 )$ plane, and the
strong contingent of maverick events all make this a tricky region of 
the MSSM IP space to analyze via the methodology adopted here.
On the other hand, a sufficiently complicated wedgebox plot may 
indicate that Nature has chosen a point in this 
relatively small (especially given the portion ruled out by LEP)
sector of the IP space.

As one moves to higher values of $\mu$ and $M_2$, the wedgebox shapes 
become much less sensitive to small shifts in the $( M_2 , \mu )$ plane.
In the region of the upper island
the wedgebox pattern almost exclusively consists of either a 
wedge, a `double-wedge,' or a box. 
This is primarily because the chargino-related production modes are
far weaker in this region, and thus the dominant source of events is
direct neutralino pair production, which as stated earlier is basically
just $\widetilde{\chi}_2^0 \widetilde{\chi}_3^0$ production, along with
Higgs-mediated neutralino pair production.
In fact at sufficiently high values of $M_2$ or $\mu$ only this latter 
contributes via $H^0/A^0 \to  \widetilde{\chi}_2^0 \widetilde{\chi}_2^0$, 
giving a simple box shape.
In the interior of the upper island
direct $\widetilde{\chi}_2^0\widetilde{\chi}_3^0$ production yields one 
wedge with an inner (outer) edge at  
$M_{i1}(\ell^+\ell^-)$ for $i=2(3)$. 
Significant Higgs-mediated modes in this region are
$H^0/A^0 \to  \widetilde{\chi}_2^0 \widetilde{\chi}_2^0$,
$\widetilde{\chi}_2^0 \widetilde{\chi}_3^0$ and
$\widetilde{\chi}_2^0 \widetilde{\chi}_4^0$. 
However, for most points on the upper island the 
crucial $\widetilde{\chi}_2^0 \widetilde{\chi}_4^0$ contribution
is too faint to give a distinct edge, and the other Higgs processes 
will simply reinforce the kinematical edges\footnote{Though the population 
structure within elements of the wedgebox plot will be altered.} from
$\widetilde{\chi}_2^0\widetilde{\chi}_3^0$ production, yielding a 
single wedge.  However near the maximum of this island at
$(M_2, \mu)=(190\, \hbox{GeV}, 280\, \hbox{GeV})$
the $\widetilde{\chi}_2^0 \widetilde{\chi}_4^0$ contribution is 
substantial and a second wedge extends out from the first with 
an inner (outer) edge at
$ M_{i1}(\ell^+\ell^-)$ for $i=2(4)$; giving a `double-wedge.'

In the special region where a double-wedge pattern is observed, one can
unambiguously identify the quantities
$M_{i1}(\ell^+\ell^-)$ ($i=2,3,4$)
which will put significant constraints on the neutralino and
physical slepton masses;
these in turn determine the MSSM IPs
$M_1$, $M_2$, $\mu$, $\tan\beta$ and $m_{\tilde{e}_{L,R}}$,
$m_{\tilde{\mu}_{L,R}}$. 
Fig.\ \ref{Dalitz-exmpls}$(b)$ shows the MC simulation of the double-wedge
at $(\mu, M_2) =  (190\, \hbox{GeV} , 280\, \hbox{GeV})$.
As both direct and Higgs-mediated modes contribute nearly 500 events
each at this point, the approximate locations of the three kinematic edges 
are easily visible.  Unfortunately, as a practical matter high rates such 
as these are indispensable in comfortably distinguishing a double-wedge 
from the more common (across the parameter space) single-wedge wedgebox 
plots.  In fact, there is no definitive division between the two wedgebox 
shapes:  quantitative criteria must be developed to gauge how many events
are required to adequately resolve the outer edge of the longer wedge.
For the moment visual inspection is all that is employed.
In this double-wedge one may by eye identify
$M_{21}(\ell^+\ell^-)=55 \pm 5\, \hbox{GeV}$,
$M_{31}(\ell^+\ell^-)=75 \pm 5\, \hbox{GeV}$,
and $M_{41}(\ell^+\ell^-)=175 \pm 10\, \hbox{GeV}$.
A more thorough analysis would of course employ a  
statistical likelihood analysis of these edges. It is already clear 
though the two-dimensional wedgebox plot offers a more  
precise method of edge-identification than the traditional
one-dimensional projection as one can, for instance, see shifts between 
the edge locations along the the axes in what would, in a one-dimensional 
plot, be taken as just one broader edge, and also discard some of the
maverick events that fall outside the regular structures of the wedgebox 
plot.  Further, anomalies in the population densities of 
structures presumed to be mirror images of each other along the two axes
may become apparent.  
For the MSSM point plotted, the actual edges, calculable from the MSSM 
IPs, are
$M_{21}(\ell^+\ell^-)=56.9\, \hbox{GeV}$,
$M_{31}(\ell^+\ell^-)=76.8\, \hbox{GeV}$,
$M_{41}(\ell^+\ell^-)=171.7\, \hbox{GeV}$.
on the muon side, and
$M_{21}(\ell^+\ell^-)=56.6\, \hbox{GeV}$,
$M_{31}(\ell^+\ell^-)=76.9\, \hbox{GeV}$,
$M_{41}(\ell^+\ell^-)=172.6\, \hbox{GeV}$ on the
electron side\footnote{This slight asymmetry arises from differences 
in the physical selectron and smuon masses which are due only to the 
fact that  $m_\mu \ne m_e$ --- here $m_{\tilde{\mu}_{L,R}}$ and
$m_{\tilde{e}_{L,R}}$ are left degenerate.  See earlier comments though
concerning the event generator.}.
Agreement with the values above obtained from visual inspection is 
reasonable.

Though sub-dominant, a significant number of the events at points on 
the upper island still do arise from chargino production. For a 
quantitative estimate, a random sampling of non-Higgs-mediated 
events passing all cuts near the maximum of the second island was 
examined: $75\%$ of these were direct neutralino production
$\widetilde{\chi}_2^0 \widetilde{\chi}_3^0$ events yielding the 
expected inner wedge, while almost all the remaining events involved
chargino production.  Approximately half of the chargino events
are `3+1' events from chargino pair production, either
$\widetilde{\chi}_2^\pm \widetilde{\chi}_1^\mp $ or
$\widetilde{\chi}_2^\pm \widetilde{\chi}_2^\mp $.  The chargino yielding
three leptons either decays via 
$\widetilde{\chi}_2^\pm \rightarrow \widetilde{\chi}_2^0 W^{\pm}$
(with the $W^{\pm}$ decaying leptonically),
reinforcing the edges of the box in the lower left corner of the wedgebox 
plot, or via
$\widetilde{\chi}_2^\pm \rightarrow Z^0 \widetilde{\chi}_1^\pm$, 
leading to $Z$-lines on the wedgebox plot.  The lighter chargino,
$\widetilde{\chi}_1^\pm$, decays over $90$\% of the time into a 
sneutrino and a single lepton, and the remaining $<10$\% of the time
into a charged slepton and a neutrino, again yielding exactly one
lepton.  
Almost all the remaining chargino events were from  
$\widetilde{\chi}_2^\pm \widetilde{\chi}_{2,4}^0$
chargino-neutralino production.  Here the $\widetilde{\chi}_2^\pm$
decays via
$\widetilde{\chi}_2^\pm \rightarrow \widetilde{\chi}_2^0 W^{\pm}$,
with the $W^{\pm}$ decaying hadronically yet not leading to jets
strong enough to violate the jet cut.

\begin{figure}[!t]
\vskip -1.0cm
\begin{center}
\includegraphics[width=4.0in]{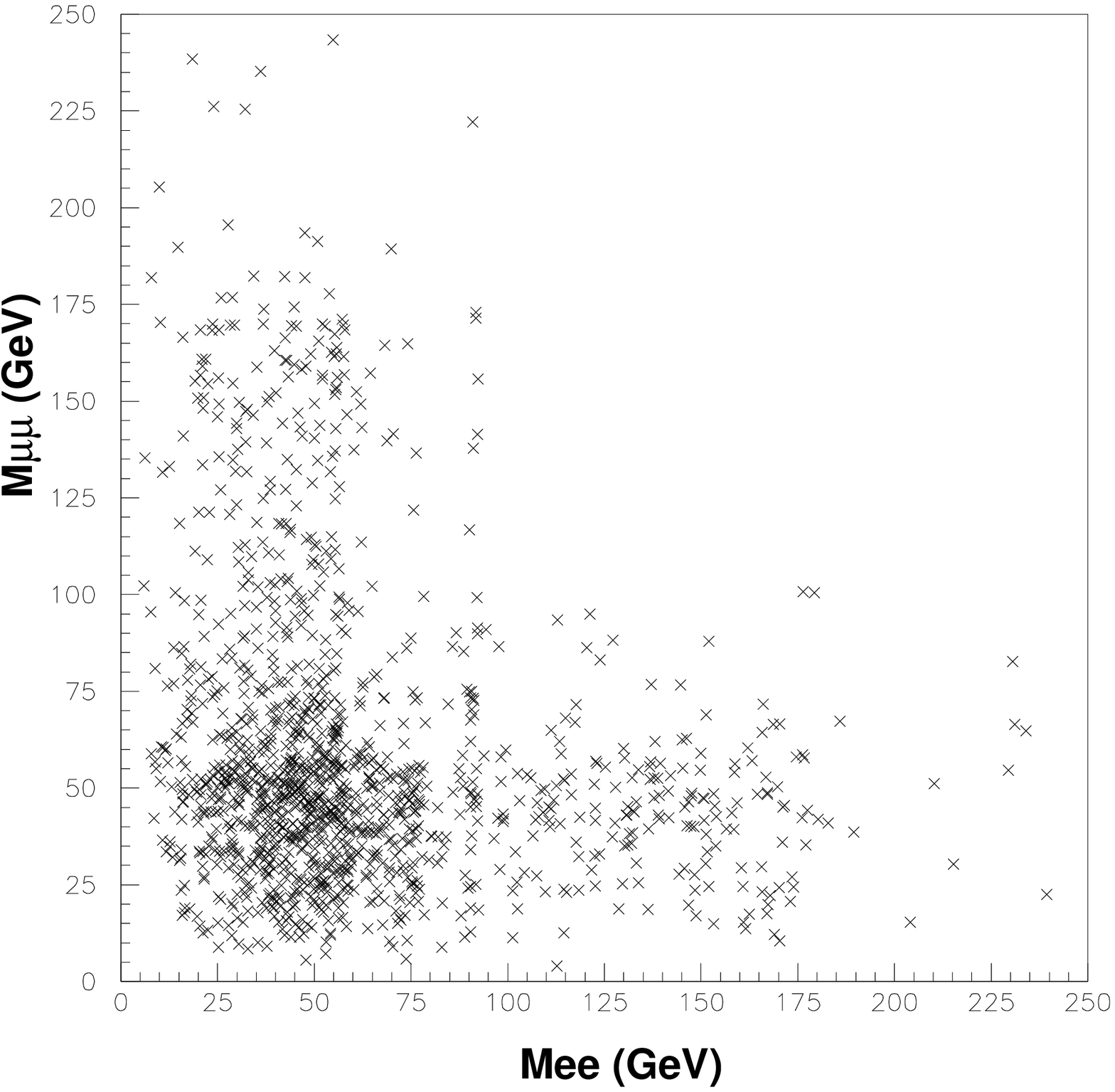}
\\
\includegraphics[width=2.0in]{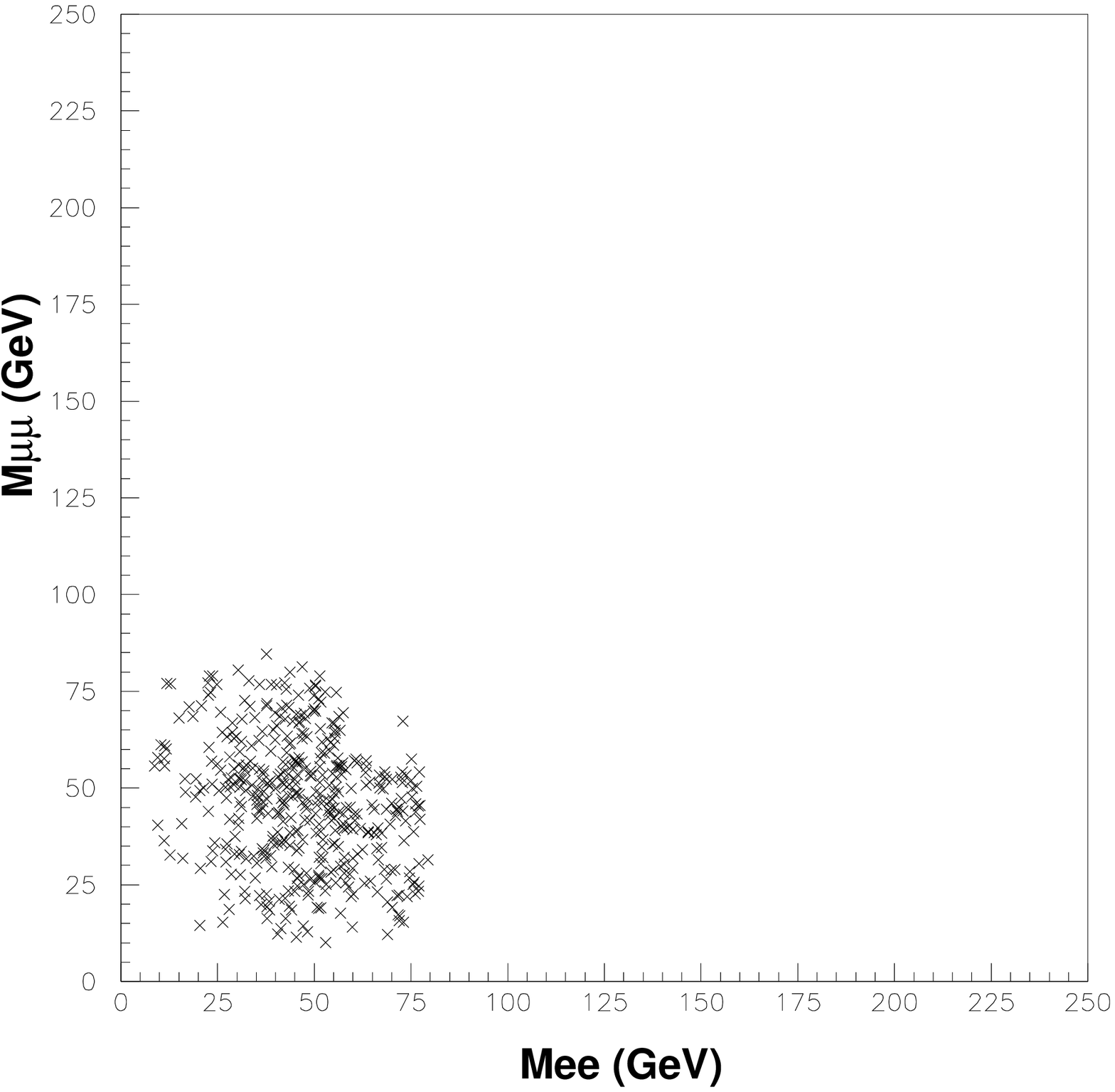}%
\includegraphics[width=2.0in]{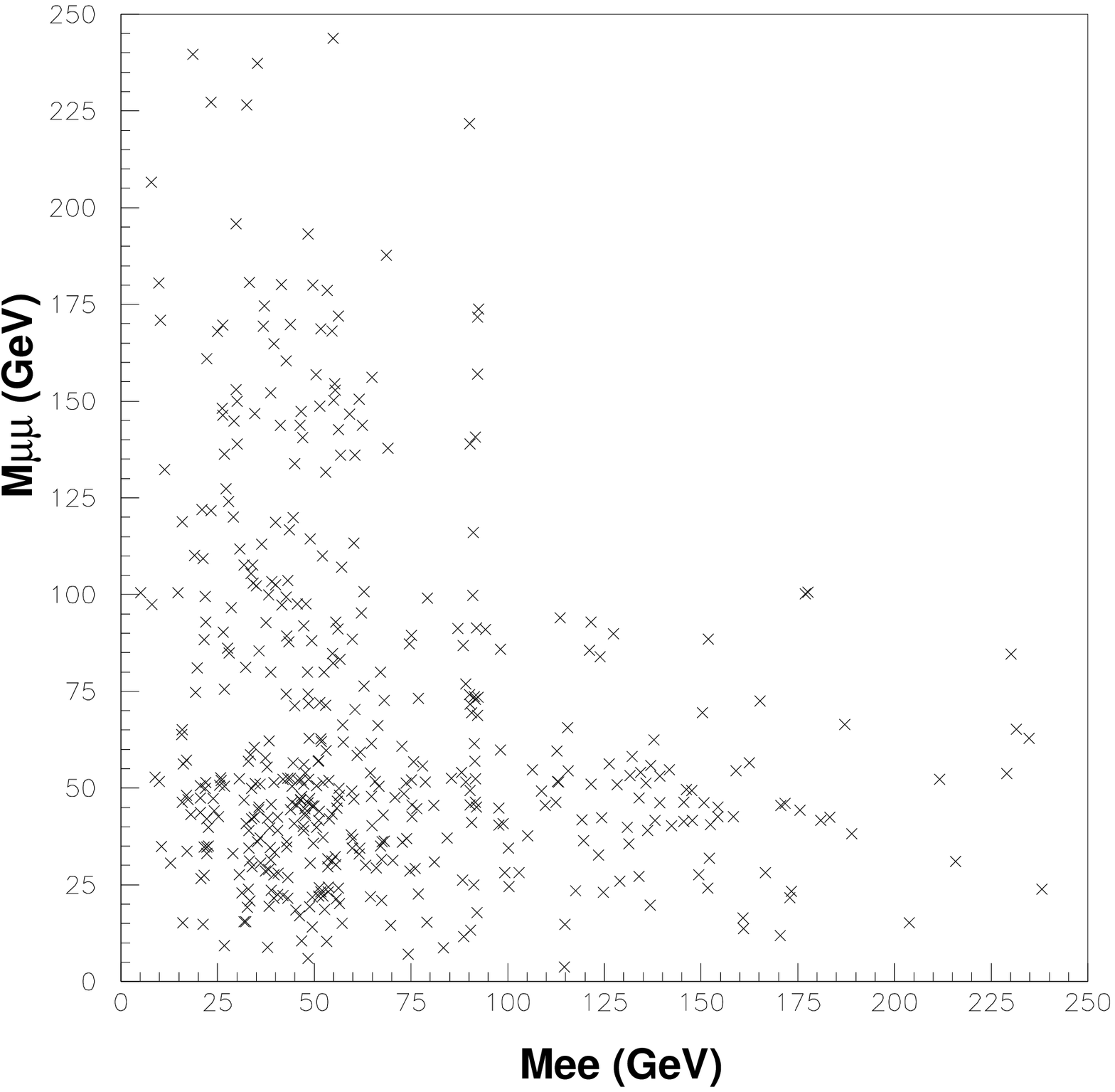}
\includegraphics[width=2.0in]{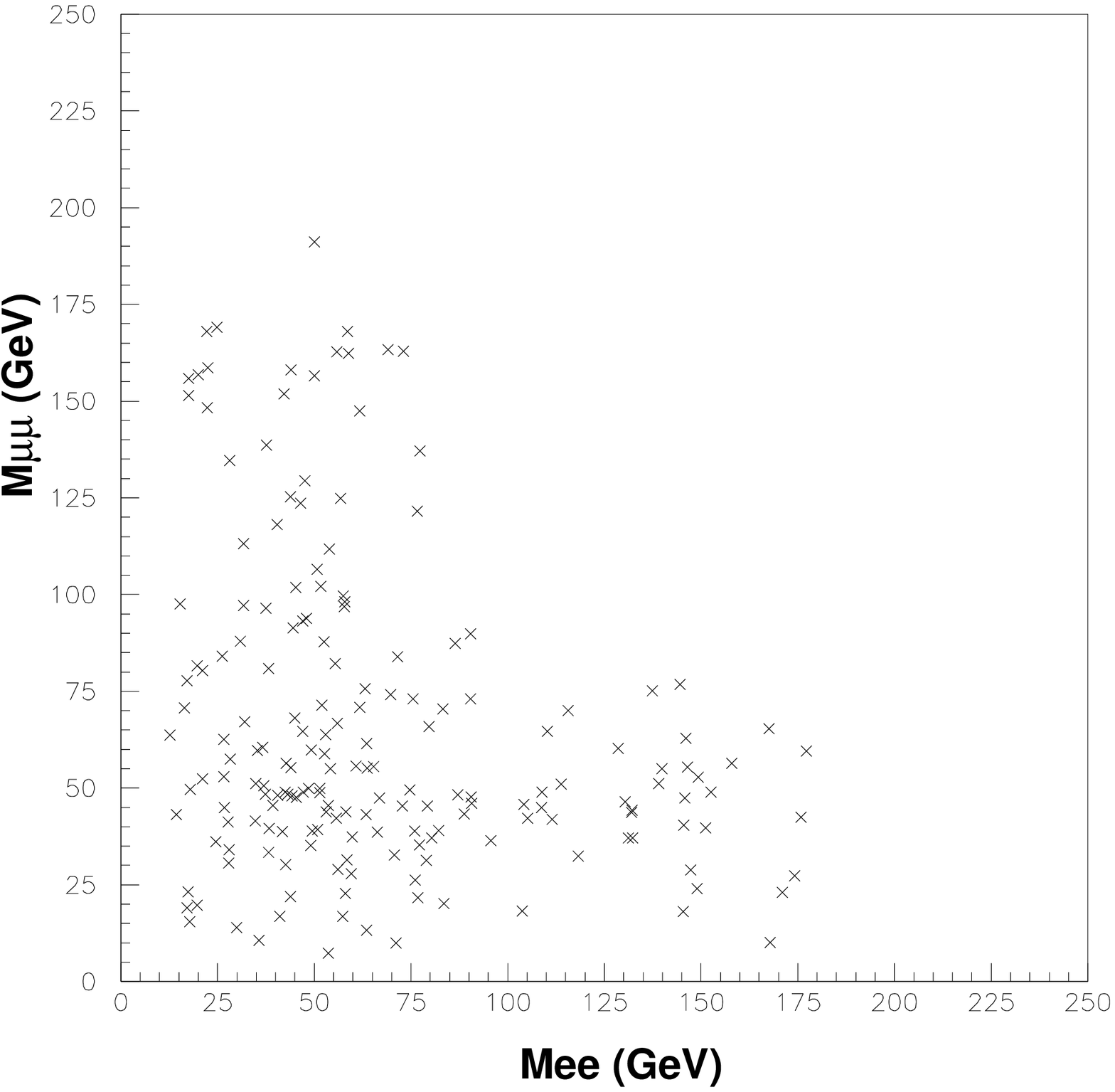}%
\includegraphics[width=2.0in]{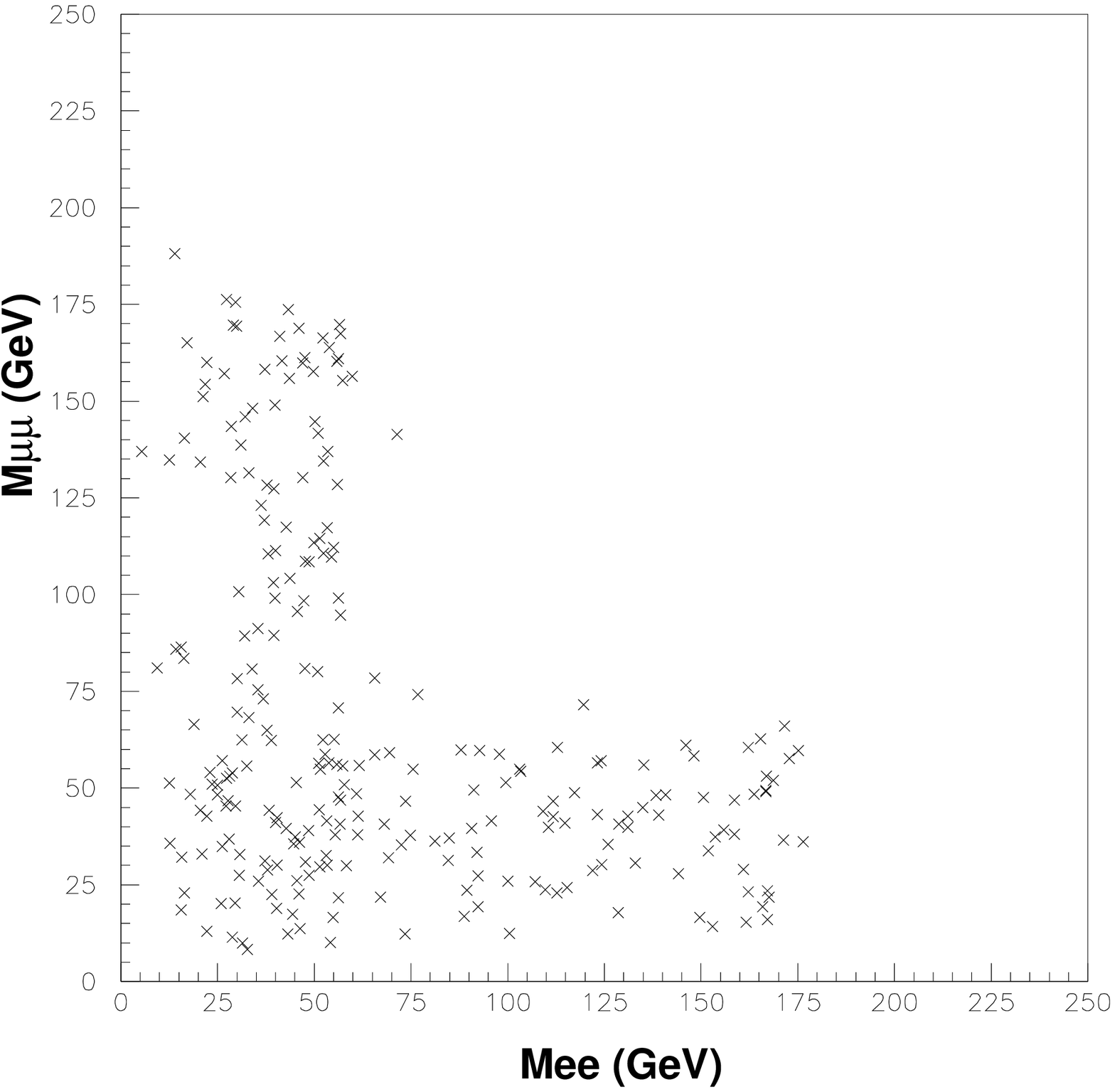}
\end{center}
\vskip -0.4cm
\caption{The wedgebox plot at $(\mu, M_2) = (190,280)$ and
$\tan\beta=10$, assuming an 
integrated luminosity of  
$300\, \hbox{fb}^{-1}$ (center top); broken down into its
$\widetilde{\chi}_2^0 \widetilde{\chi}_3^0$ component (upper left),
 $\widetilde{\chi}_i^\pm\widetilde{\chi}_j^\pm + 
\widetilde{\chi}_i^\pm \widetilde{\chi}_i^0$ component (upper right),
and $H^0,A^0$ components (lower left, right, respectively).   
Combining the components yields the double-wedge wedgebox plot
(`decorated' by a halo and $Z$-lines from the chargino component)
seen at the top.
Note:  this analysis was done using ISAJET \cite{ISAJET}.
 \label{separatedParts} }
\vskip -1.0cm
\end{figure}

Fig.\ \ref{separatedParts} (upper plot) shows the wedgebox plot for 
$(\mu, M_2) = (190\, \hbox{GeV}, 280\, \hbox{GeV})$, but this time 
with $\tan\beta = 10$.  Note that the result is very similar to 
that with $\tan\beta = 20$ seen in Fig.\ \ref{Dalitz-exmpls}$(b)$.
In addition, Fig.\ \ref{separatedParts} was made using the 
ISAJET 7.64 \cite{ISAJET} event generator (coupled with a basic
LHC detector simulation routine).  Using ISAJET facilitates 
separating out the contributions from each production process.
For this point in MSSM IP space, these are shown in the lower
four plots of Fig.\ \ref{separatedParts}:  the upper left plot is 
only from direct $\widetilde{\chi}_2^0 \widetilde{\chi}_3^0$
production, the upper right plot is from chargino production
processes, and the lower left (right) plot is due to $H^0$ ($A^0$)
production and subsequent decays.   Halo events, as well as $Z$-lines, 
are clearly seen to come from the `maverick' (`3+1') chargino events.
Whereas $\widetilde{\chi}_2^0 \widetilde{\chi}_3^0$ production yields a
clean, relatively short, wedge with an outer edge at 
$\sim$$80\, \hbox{GeV}$, and $H^0,A^0$ production gives a 
longer ($\widetilde{\chi}_2^0 \widetilde{\chi}_4^0$) wedge
terminating around $175$-$180\, \hbox{GeV}$.   A population
change at around $75$-$80\, \hbox{GeV}$ is also marginally discernible 
for the Higgs-mediated components.
Note carefully that, although the chargino production plot is far from 
structureless, the edges of the (vaguely) wedge-like structure seen
in this plot (a noticeable drop in the event population is apparent in the 
vicinity of $175\, \hbox{GeV}$) merely reinforce the edges seen in the 
$H^0,A^0$ plots (and perhaps the 
$\widetilde{\chi}_2^0 \widetilde{\chi}_3^0$ plot as well), while 
the halo events and $Z$-line events do not prevent one from identifying
the $\widetilde{\chi}_2^0 \widetilde{\chi}_3^0$ and $H^0,A^0$ edges.
This is true of all the upper island wedgebox plots examined:  the 
maverick event characteristics do not obscure identification of the 
sought-after neutralino-based endpoints.  The chargino-derived $Z$-line 
events do make the overall wedgebox plot look fatter, and one must be 
cautious not to be misled by mavericks which partially fill in the space
between the $Z$ lines and the wedges.
Note that only the chargino-related component generates $Z$-lines.
Adding effects from all these components produces a fairly 
easily-recognizable double-wedge.

As noted earlier, Fig.\ \ref{wedgemap} is thus a
representative example of a class of $(\mu , M_2)$ wedgebox plots 
for one set of $m_A$, $\tan\beta$ and slepton inputs.  Now to examine
variation within this class, first consider lowering 
$\tan\beta$ from $\tan\beta = 20$:  the total 
rate drops as seen in Fig.\ \ref{MCvartanB}. However if a $30$ event 
contour were added to each of the plots in  Fig.\ \ref{MCvartanB}, it 
would actually cover the entire region shown; thus, for an integrated 
luminosity of $300\, \hbox{fb}^{-1}$ at the LHC, at least ${\sim}100$ 
events may be collected at all points shown for each $\tan\beta$ value, 
sufficient to build a respectable wedgebox plot. 
Since it is really the Higgs boson processes which are losing rate while 
other processes are roughly constant, we expect wedgebox shapes at each 
point to remain more-or-less the same: on the lower island the minor Higgs 
boson contribution does not markedly affect the already complicated 
chargino-dominated wedgebox structure, 
while on the upper island removing the Higgs contribution at worst removes 
the edge associated with 
$H/A \to  \widetilde{\chi}_2^0 \widetilde{\chi}_4^0$
(since $H^0/A^0 \to  \widetilde{\chi}_2^0 \widetilde{\chi}_{2,3}^0$ simply 
reinforce edges already present from direct- and chargino-channels). 
So for lower $\tan\beta$ values, it may not be possible to observe the 
double-wedge:  somewhere between $\tan\beta=10$
and  $\tan\beta=20$ the '300' event (per $100\, \hbox{fb}^{-1}$) contour 
vanishes, meaning it becomes questionable whether or not there is enough
statistics to clearly identify the outer wedge of the double-wedge.  
Thus analogous plots to Fig.\ \ref{wedgemap} for $\tan\beta$ values of 
$5$ or $10$ look quite similar to the Fig.\ \ref{wedgemap} save that
more and more double-wedge wedgebox plots will become simple single-wedge
wedgebox plots.
As $\tan\beta$ rises above $20$, rates may continue to rise --- if 
SUSY-breaking stau mass inputs are set safely above the corresponding 
selectron and smuon inputs, as has been done by hand here ---  or they
may plummet (as noted earlier) --- if stau inputs are made degenerate with 
those of the first two generations, in which case mixing will increasingly 
drive down one of the physical stau masses as $\tan\beta$ grows, cutting 
down 
`leptonic' (that is, electron and muon) BRs in favor of decays yielding taus.

The preceding paragraph touched upon the dependence of the appearance of 
Fig.\ \ref{wedgemap} on inputs of the slepton sector, specifically, the 
value(s) of the stau inputs relative to those of the first two 
generations.  Now, setting aside the staus for the moment, consider how
changing the selectron and smuon inputs will affect results.
If the first two generations' degenerate input mass is raised above 
(lowered below) the nominal value of $150\, \hbox{GeV}$ assumed in 
Fig.\ \ref{wedgemap}, all rates decline (grow).  Repeating here the rough 
estimates given above, if rates fall significantly below $1000$ events per 
$300\, \hbox{fb}^{-1}$ at an upper island point in the MSSM IP space, 
double-wedge wedgebox plots tend to become single-wedge wedgebox plots, 
and if rates drop much below $100$ events per $300\, \hbox{fb}^{-1}$, 
no clear wedgebox pattern may emerge at all.  Thus by merely determining 
the overall event rate at a point and then referencing 
Fig.\ \ref{wedgemap}, a good estimate of what the wedgebox plot at that 
point should look like can often be deduced.  If the degeneracy of the 
slepton mass inputs is lifted, then rates can be raised even if the 
left slepton inputs are made more massive, since the sneutrino masses
also rise with the left slepton inputs, choking off the spoiler modes
(see Fig.\ \ref{MCslepvar}).  If smuon SUSY-breaking mass inputs are 
pushed above those of the selectrons, rates may drop if significant
non-sleptonic neutralino decay modes exist.  In addition, the widths of 
the kinematical edges would shrink (anywhere from very modestly to 
enormously).

Lastly, consider the impact of altering $m_A$.  This would affect the 
Higgs boson contributions (the focus of \cite{ha4l}) to the wedgebox plot.  
If $m_A$ is lowered significantly, then the only open Higgs decay channel 
to neutralinos would be to $\widetilde{\chi}_2^0 \widetilde{\chi}_2^0$.
This would produce a $\widetilde{\chi}_2^0 \widetilde{\chi}_2^0$ box.
If direct $\widetilde{\chi}_2^0 \widetilde{\chi}_3^0$ production is
significant, then the Higgs boson-induced 
$\widetilde{\chi}_2^0 \widetilde{\chi}_2^0$ box would lie at the corner of 
the $\widetilde{\chi}_2^0 \widetilde{\chi}_3^0$ wedge, producing no new 
edges and thus indistinguishable {\em by shape} from a wedgebox plot 
consisting solely of a $\widetilde{\chi}_2^0 \widetilde{\chi}_3^0$ wedge.
However, the presence of the  Higgs boson-induced
$\widetilde{\chi}_2^0 \widetilde{\chi}_2^0$ box may well be noticeable via 
the population structure of the various component parts of the wedge ---
the Higgs boson decays in this case will over-populate the corner box of 
the wedge.  The Higgs boson is critical to producing the very desirable
double-wedge pattern, and, to allow this, $m_A$ must be large enough to 
allow $H^0$ and\footnote{For larger values of $m_A$ relevant here,
$H^0$ and $A^0$ are nearly degenerate.} $A^0$ to decay to 
$\widetilde{\chi}_2^0 \widetilde{\chi}_4^0$.  Yet if $m_A$ is made too 
large, Higgs boson production rates drop off and the Higgs boson 
contribution dies.   The event number estimate stated in the previous 
paragraph again can serve as a guide as to when clear double-wedge
wedgebox plots may be identified.

\section{Discussion}
\label{sec:disc}

\begin{figure}[!htb]
\vskip -1.0cm
\begin{center}
\dofig{4.10in}{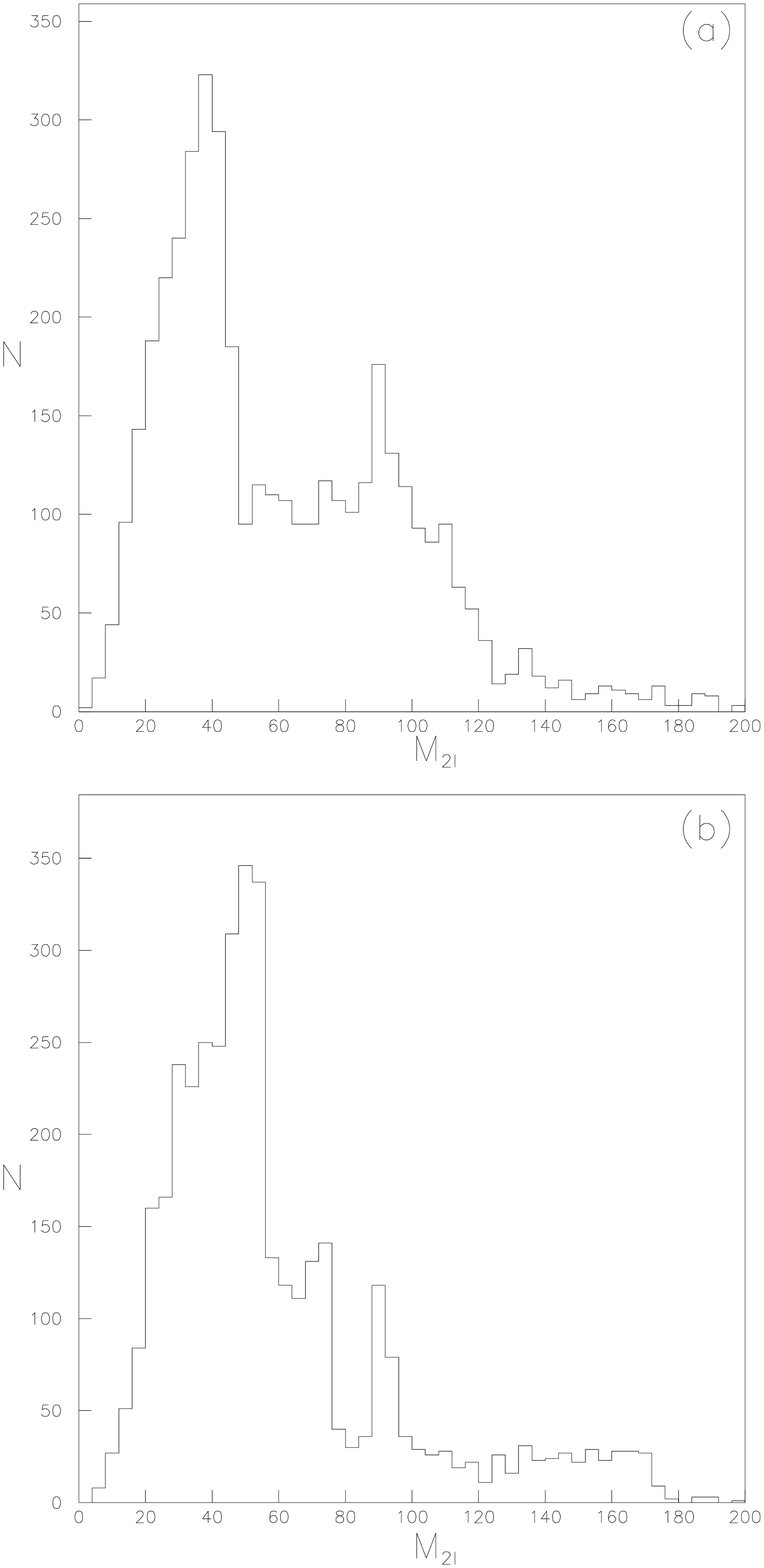}
\end{center}
\vskip -0.6cm
\caption{ 1-dimensional projections of the two wedgebox plots in
Fig.\ \ref{Dalitz-exmpls} obtained by putting values of both
$M(e^+e^-)$ and $M({\mu}^+{\mu}^-)$ for each event on one axis.
 \label{1dim} }
\end{figure}

Fig.\  \ref{1dim} shows the
conventional one-dimensional projections which may be obtained
from the two wedgebox plots in Fig.\  \ref{Dalitz-exmpls} by 
plotting both the
$M(e^+e^-)$ and the $M({\mu}^+ {\mu}^-)$ values from each event
along a single axis.
While mass differences may still be inferred from sharp changes in
curve (though care must be taken in interpreting, for instance, the 
maverick-induced glitch near $60\, \hbox{GeV}$ in the upper plot
or rate increases around the $Z^0$ pole in both plots), information 
gained from correlating the $M(e^+e^-)$ and $M({\mu}^+ {\mu}^-)$ values 
is clearly lost, meaning that whether the events are generated chiefly 
by similar or dissimilar neutralino pair production may no longer be 
determined.  Further, the ability to identify and thus exclude 
so-called `maverick' events outside of the wedge and box geometrical
elements may significantly increase the purity of the sample events and 
the resulting resolution of the kinematical edges.

Two-dimensional wedgebox plots contain considerably more 
information often packaged in a manner readily understandable. 
For instance, a double-wedge or wedge-protruding-from-a-box wedgebox plot 
almost always\footnote{This general rule can be broken by 
$\widetilde{\chi}_2^{\pm} \to \widetilde{\chi}_1^{\pm} \ell^+ \ell^-$ 
decay events or by so-called `stripes' (see \cite{cascade}) where
$\widetilde{\chi}_i^0 \to \widetilde{\chi}_j^0  \ell^+ \ell^-$ 
($j \ne 1$).  The former was mainly encountered in the region of the 
lower island and the latter was never found to be significant.}
has edges corresponding to $\widetilde{\chi}_i^0 \to \widetilde{\chi}_1^0$ 
decays ($i=2,3,4$ in order proceeding out along either axis).  Though 
these values may not equal the neutralino mass differences, if the 
intermediate slepton or $Z^0$ is on mass-shell 
({\it cf.}, Eqn. \ref{edge2body}), in any case much of information 
concerning the inputs to the MSSM neutralino mixing matrix and slepton sector 
may certainly be obtained.  Technically, in other wedgebox patterns with 
fewer edges there is an inherent ambiguity in identifying the 
$\widetilde{\chi}_i^0 \to \widetilde{\chi}_1^0$ decay responsible for each 
edge, though in practice taking the lowest available $i$ for each edge 
is usually the correct choice; and the information concerning the MSSM
input parameters extractable remains substantial.

Furthermore, commensurate with being able to compartmentalize the 
two-dimen-
\newline
sional space of a wedgebox plot into a collection of simple 
geometrical shapes is the ability to examine the population densities 
in each element --- for instance
how many events would populate one leg of a wedge as compared to  
a box on the same wedgebox plot?  Herein lies a true advantage of
the wedgebox plot over the 1-dimensional projection.  Theoretically, 
the distribution of the population of events within a given element 
is expected to be fairly simple, as noted in the introduction, at 
least before the implementation of cuts.  Thus the expected the number of 
events in the corner box of a wedge versus the number in the legs is 
reasonably easy to estimate.
This specifically allows us to tell whether Higgs boson 
$H/A \to \widetilde{\chi}_2^0\widetilde{\chi}_2^0$ production is
present on top of a direct
$\widetilde{\chi}_2^0\widetilde{\chi}_3^0$ channel-dominated 
wedge by virtue of the overpopulated corner box.
Note this information is much harder to extract from 
a 1-dimensional projection.
 
In other analyses examining signals for the heavier MSSM Higgs bosons
\cite{ha4l, charg2} one typically selects a point in the MSSM IP space 
{\em and then } computes the signal rate from the Higgs boson(s)
and the background rates from other MSSM processes\footnote{Of course
the SM backgrounds must also be considered, but these do not vary 
with the MSSM IPs.}.  Then if a large enough excess from the Higgs
boson `signal' is seen over the MSSM (+SM) `background' a discovery
or detection of the Higgs boson can be projected {\em at this point 
in the MSSM IP space}.  But this raises the question, could the 
excess events attributed to the Higgs bosons at one point in 
the MSSM IP space be swallowed up by a larger SUSY background 
rate at another point in the MSSM IP space?   This question is 
virtually never addressed, since in these studies 
it would be much too computationally impractical to find the 
background rates at an infinite number of points spanning the whole 
of MSSM IP space.

Contrast this with what may be inferred from the preceding studies 
connected with Fig.\ \ref{wedgemap} and the
variation of the MSSM IP parameters fixed to generate this figure.
Said studies indicate that all wedgebox maps should look similar to 
Fig.\ \ref{wedgemap}, except the distance from the islands' maxima
where the number of events becomes too small to clearly identify any
wedgebox pattern changes and, on the upper island\footnote{The lower 
island is a domain of rapidly varying (as one moves around the IP space)
wedgebox patterns which tend to be fairly complicated.  This complexity 
serves to pinpoint the location in the parameter space as being on this 
lower island.}, double-wedge wedgebox
plots shift into single-wedge wedgebox plots (or {\it vice versa}).  The
steadiness of the features in Fig.\ \ref{wedgemap} allow 
several fairly robust conclusions to be drawn, within the context of the 
MSSM:
\begin{itemize}
\item  any box-containing wedgebox
plot (including patterns with an outer box envelope,
patterns with a wedge protruding from a box (as in 
Fig.\ \ref{Dalitz-exmpls}$(a)$), and patterns where a box is 
inferred through the over-population of the corner of a wedge) 
indicates that either Nature sits on the lower island in the lower
corner of the $(\mu, M_2)$ plane or events from $H^0$ and $A^0$
decays are present and substantial.  The key here is the establishment 
that substantial direct neutralino pair production can only occur
for $\widetilde{\chi}_2^0\widetilde{\chi}_3^0$ production, which yields
a wedge not a box.  Also it is crucial to exclude cascade processes via 
a sufficient cut on hadronic activity.

\item The severity of the halo events, which stem from chargino and
slepton production processes (especially `3+1' events), around the 
wedge and box geometries expected in the wedgebox theoretical framework
indicates how near Nature lies to one of the axes.  High levels of 
such contamination are found near the axes,  for low $\mu$ and/or
$M_2$ values, while, conversely, very `clean' wedge like plots 
indicate moderate values of
$\mu , M_2$ ($\sim 200 \, \hbox{GeV}$) and direct channel 
$\widetilde{\chi}_2^0\widetilde{\chi}_3^0$ domination.

\item With rare exceptions, a double-wedge wedgebox plot 
unambiguously identifies three
kinematical endpoints $M_{i1}(\ell^+\ell^-)$ ($i=2,3,4$) of
neutralinos decaying through off-shell sleptons or $Z^{0*}$.

\end{itemize}

Ideally, one would like to make even stronger statements along the lines
that if one sees a certain wedgebox pattern, this unambiguously means one 
is seeing evidence for the heavier MSSM Higgs bosons, regardless of the 
specific point in the MSSM IP space Nature has chosen.  While the present 
analysis does not quite reach this goal, it is reasonable to expect that 
more detailed criteria can lead to definite conclusions concerning such 
issues.  Nonetheless, the conclusions that may be drawn from the simple 
gross properties of wedgebox plots as herein presented is most encouraging.  
With the choices of where Nature might lie in the MSSM IP space narrowed 
down by such an analysis, the full weight of more intricate probing of the 
data (via neural network studies and their kin for instance) can optimize
(though perhaps at the expense of clarity in the presentation of these 
results to those not immersed in the intricate details of hadron collider
analyses of SUSY phenomenology) the amount of information extractable at 
the LHC.

Another handle that may aid in determining if heavy MSSM Higgs bosons 
are generating some of the $e^+e^- \mu^+\mu^- \, + \, \slashchar{E}$
events is the invariant mass of all four leptons combined.  This 
is expected to be bounded above by 
$m_{H,A} - 2m_{\widetilde{\chi}_1^0}$ \cite{Hawaii}.  However, the
four-lepton (one-dimensional) invariant mass distribution will not have 
the abrupt turn off expected for the di-lepton invariant masses plotted 
in the wedgebox plot \cite{Hawaii}.  Studies suggest that the backgrounds 
and the low number of Higgs boson-generated events near the endpoint are 
likely to obscure detection of the endpoint.  However, the shape of a 
histogram plotting the four-lepton invariant mass distribution may be 
markedly affected by having a significant fraction of the events coming 
from Higgs boson decays.  Drawing conclusion from the distribution shape
also encounters problems though since the shape of the MSSM background
(as well as that of the Higgs boson signal) varies across the MSSM IP
space, again leading to the unattractive methodology of first picking a 
point in the parameter space and then determining if the signal $+$ 
background distribution differs significantly from the background alone
distribution {\em at this point and this point alone} in the MSSM IP 
space.  Conclusions drawn from the wedgebox plot may help to alleviate
some of this uncertainty, enabling the four-lepton invariant mass 
distribution to be more successfully employed.

\section{Conclusions}
\label{sec:conc}

The wedgebox technique may be used to effectively and elegantly categorize 
any positive outcome of an LHC search for the 
$e^+e^- \mu^+\mu^- \, + \, \slashchar{E}$ signature expected from 
MSSM neutralino pair production.  A search of the entire available 
MSSM IP space reveals that a sufficient number of events to make a 
viable wedgebox plot (somewhat arbitrarily set as being $> 100$
assuming an integrated luminosity of $100\, \hbox{fb}^{-1}$)
is obtained only on two `islands' in the 
$(\mu , M_2)$ plane (as shown in Fig.\ \ref{MCvartanB}).  Here it is 
assumed slepton masses are relatively low with
$m_{\tilde{\ell}_L} \simeq m_{\tilde{\ell}_R}$ --- if this equality
is altered, then the strengthening (weakening) of sneutrino spoiler modes
tends to make the islands sink (the moat between the islands dry up) for 
$m_{\tilde{\ell}_L} <$ ($>$) $m_{\tilde{\ell}_R}$.

Much of the lower island is already excluded by negative search results at 
LEP.  Signature events on the small as-of-yet unexcluded portion of this 
island result from a smorgasbord of different production processes
including a large, even dominant, component from processes involving
charginos.  Not unexpectedly, a plethora of wedgebox plot patterns 
result, with one pattern shifting into another fairly rapidly as the 
exact location in the MSSM IP space shifts.  A weakness of the 
wedgebox plot technique is that it does not fully include charginos 
into its theoretical framework (at least not thus far).  What can be
said is that if a wedgebox plot with a complicated structure is observed, 
then this points toward Nature resting on this small portion of this 
lower island.  It would be quite a coup if this were observed, and would 
no doubt motivate more detailed examinations to pry more information 
from the rich though complicated characteristics of this MSSM IP region.

The last statement is possible because of the simple character of the larger
upper island: here the wedgebox pattern is remarkably constant, consisting 
either of a single- or double-wedge.
This is mostly due to the fact that the only neutralino pair to be 
directly produced at any
appreciable rate is $\widetilde{\chi}_2^0 \widetilde{\chi}_3^0$, and this 
yields a wedge pattern.
Furthermore, unlike on the lower island, chargino 
production processes on the upper island are of a more tame variety,
and mostly fortify the wedge structure (though halo events and sometimes
$Z$-lines are added to the underlying wedgebox structure one hopes to 
categorize), while slepton backgrounds are negligible.
The clarity of the double-wedge pattern on the upper island depends
on whether Higgs-mediated neutralino pair production 
yields a sufficient number of $\widetilde{\chi}_2^0 \widetilde{\chi}_4^0$
events for detection of the kinematical edge resulting from 
$\widetilde{\chi}_4^0 \to \widetilde{\chi}_1^0 \ell^+\ell^- $
decays.
Therefore if the MSSM IPs are tuned to give a very clear double-wedge
pattern ({\it i.e.}, at the level of Fig.\ \ref{Dalitz-exmpls}$(b)$ or 
better), or a wedge-protruding-from a box pattern, then one can directly 
read off the kinematic edges $M_{i1}(\ell^+ \ell^-)$, ($i=2,3,4$), 
which strongly constrain the neutralino and slepton spectra and the 
corresponding MSSM IPs.

Another fairly sweeping general result emerges from studying the variation 
of the wedgebox plot patterns across the MSSM IP space:
the presence of a box in a wedgebox plot, where hadronically noisy
events from gluino and squark cascade decays have been removed,
signals either chargino production or 
heavy Higgs boson-mediated neutralino pair production, where 
the former only generates suitably-resolved boxes in the quite restricted 
region of the MSSM IP space around $\mu, M_2 \, \lsim \, 200\, \hbox{GeV}$ 
(the location of the lower island).
Here boxes include box-like outer envelopes to the wedgebox pattern,
boxes with wedges protruding from them, and boxes identified via 
over-populated lower-left corners of wedges. 
Compare this result to analyses that attempt to prove the presence of 
heavy Higgs bosons by looking for excesses in the number of expected 
background events from SM {\em and other MSSM processes} on the basis 
of point-by-point studies in the MSSM IP space.  What, other than further 
{\em typically unspecified} analyses, is to say that the excess 
attributed to Higgs bosons at one point studied in the IP space could not 
be due to larger background MSSM process rates at some other 
unstudied point in the IP space?

We note the above conclusions are justifiable only by the ability to 
distinguish correlations between $M(e^+ e^-)$ and  $M(\mu^+ \mu^-)$ in the 
wedgebox technique\footnote{It also appears possible \cite{EEEEMUMUMUMU} 
to use $e^+ e^-e^+ e^-$ and $\mu^+ \mu^-\mu^+ \mu^-$ events and use 
relatively simple criteria to correctly (a high percentage of the time) 
pair up the leptons.  This will approximately double the event rates.}, 
this being manifestly impossible in the more traditional 1-dimensional 
invariant mass histograms like those shown in Fig.\ \ref{1dim}. 
Other advantages of this technique include 
(1) there is a one-to-one correspondence between an four-lepton event
and a point on the plot, (2) asymmetries between slepton generations can be
observed, and (3) better resolution of kinematic edges is possible 
by way of cutting out maverick and Z-line events which protrude from the 
dominant wedgebox shape. 

The method therefore represents a technical improvement.
Though it must be cautioned that this does not imply that the 
technique is guaranteed to unambiguously determine the presence of the 
heavier MSSM Higgs bosons, or yield a double-wedge wedgebox pattern from 
which loads of information on MSSM IPs can be mined, or restrict said 
MSSM IPs to the lower island by the complexity of the observed wedgebox 
plot.  Regrettably, at many points in the MSSM IP space, the 
$e^+e^- \mu^+\mu^- \, + \, \slashchar{E}$ signal rate is too low to 
construct a wedgebox plot (see footnote 16 for a partial remedy for 
this though).  It is perhaps also useful to bear in mind that Nature will 
select out one and only one point in the MSSM IP space.  And Nature may 
not chose to pay any attention to statements about what is true over
large stretches of the IP space in making said selection.

Finally, note that this technique can be used to compare outputs expected 
in various different sub-spaces of the MSSM IP space, such as those 
resulting from specifying different SUSY-breaking mechanisms 
({\it e.g.}, mSUGRA or GMSB) and associated higher energy-scale features 
of the model.  It can also be applied to the NMSSM, which 
adds a fifth neutralino, to little Higgs (extra-dimensions) models
with T-parity \cite{TP} (KK-parity \cite{KK}),
or indeed (as noted in \cite{cascade}) to any model in which heavy exotic 
particles $X_i$ decay to dilepton pairs $X_i \to \ell^- \ell^+$ and there 
is more than one $X_i$. To ascertain the potential usefulness of the 
wedgebox technique in such models a wedgebox map or maps may be 
constructed covering the relevant parameter space.

\end{document}